\documentclass[twocolumn,showpacs,amsmath,amssymb,superscriptaddress]{revtex4-1}
\usepackage{dcolumn}
\usepackage{color}
\usepackage{graphicx}
\usepackage{amsmath}
\usepackage{multirow}
\usepackage{bm}
\usepackage{url}

\newcommand\+{\dagger}

\begin{document}

\title{
Signatures of octupole correlations in neutron-rich odd-mass barium isotopes
}

\author{K.~Nomura}
\affiliation{Physics Department, Faculty of Science, University of Zagreb, 10000 Zagreb, Croatia}

\author{T.~Nik\v si\'c}
\author{D.~Vretenar}
\affiliation{Physics Department, Faculty of Science, University of Zagreb, 10000 Zagreb, Croatia}

\date{\today}

\begin{abstract}

Octupole deformation and the relevant spectroscopic properties of
 neutron-rich odd-mass barium isotopes are investigated in a theoretical 
 framework based on nuclear density functional theory and the particle-core
 coupling scheme. 
The interacting-boson Hamiltonian that describes the octupole-deformed
even-even core nucleus, as well as the single-particle energies and
occupation probabilities of an unpaired nucleon, are completely
determined by microscopic axially-symmetric $(\beta_{2},\beta_{3})$-deformation
 constrained self-consistent mean-field calculations for a specific choice of the energy
 density functional and pairing interaction. A boson-fermion
 interaction that involves both quadrupole and octupole degrees of
 freedom is introduced, and their strength parameters are determined to
 reproduce selected spectroscopic data for the odd-mass nuclei. 
The model reproduces recent experimental results both for the even-even and odd-mass Ba isotopes. In particular, 
 for $^{145,147}$Ba our results indicate, in agreement with recent data,
 that octupole deformation does not determine  
 the structure of the lowest states in the vicinity of the ground state, and only becomes relevant at higher
 excitation energies.

\end{abstract}


\maketitle

\section{Introduction}

Experiments using radioactive ion-beams (RIB) provide access to 
previously unknown short-lived nuclei far from the valley of $\beta$-stability.
Among the most basic nuclear properties that have been extensively explored over many 
decades are geometrical shapes of nucleon density distributions and the corresponding excitation patterns. 
In particular, reflection-asymmetric octupole (or pear-shaped) deformation
\cite{butler1996} has been a recurrent theme of interest in nuclear
structure physics, both from the  
experimental and theoretical point of view. 
Atoms with octupole-deformed nuclei are particularly relevant in the context of 
the violation of time-reversal (T) invariance in the Standard model of particle physics 
\cite{haxton1983,dobaczewski2005}.
A number of new experiments are either running or being planned 
at major RIB facilities around the world. Recently, experimental
evidence for permanent octupole deformation in radioactive nuclei has
been reported, e.g., $^{224}$Ra and $^{220}$Rn at CERN
\cite{gaffney2013} and $^{144}$Ba \cite{bucher2016} and $^{146}$Ba
\cite{bucher2017} at Argonne National Laboratory. 

Octupole deformation is also relevant for nuclei
with odd nucleon number, particularly the interplay between single-particle
degrees of freedom and the quadrupole and octupole
collective degrees of freedom that determines the structure of low-lying states. 
Furthermore, in odd-$A$ octupole deformed nuclei the Schiff moment, that is, 
a measure of nuclear time-reversal violation, is particularly enhanced. Examples 
are  $^{225}$Ra \cite{dobaczewski2005,parker2015} and $^{199}$Hg \cite{griffis1983}. 
As a wealth of new data on odd-nucleon systems become
accessible, timely and accurate theoretical studies of their
spectroscopic properties are needed. 
From the computational point of view, however, a microscopic description
of odd-mass nuclei is highly demanding, particularly in medium- and
heavy-mass systems. The reason is partly because of the fact that in the
odd-nucleon systems one must explicitly consider  
single-particle degrees of freedom, and treat them on the same level with 
collective degrees of freedom. Modelling the structure of odd-A nuclei is 
even more complicated when octupole collective degrees of
freedom are involved.

Recently we have developed a novel method \cite{nomura2016odd} for
calculating spectroscopic properties of odd-mass nuclear systems, based on
the framework of nuclear density functional theory and the particle-core
coupling scheme. 
Our approach enables an accurate, systematic and computationally 
feasible description of the odd-mass nuclei. 
The even-even core nucleus is described in the language
of the interacting boson model (IBM) \cite{IBM}, and the particle-core coupling is
taken from the interacting boson-fermion model (IBFM) \cite{IBFM}. 
The deformation energy surface for the even-even-core nucleus, the 
single-particle energies and occupation probabilities of the unpaired
nucleon are determined by a microscopic self-consistent mean-field (SCMF)
calculation
for a given energy density functional and pairing interaction. 
These mean-field results are then used as input for the IBFM Hamiltonian. 
However, several strength parameters of the
boson-fermion interaction need to be specifically adjusted to reproduce
selected data for the low-energy excitation spectra in the odd-mass nuclei.

In this work we extend the approach of Ref.~\cite{nomura2016odd} to a 
description of spectroscopic properties of odd-mass 
systems in which octupole deformation is expected to play a role. 
Here the major development of the method is that the boson-core
Hamiltonian is not only built from  
the usual positive-parity $J^{\pi}=0^+$ monopole ($s$) and $J^{\pi}=2^+$ quadrupole
($d$) bosons, but also contains the negative-parity $J^{\pi}=3^-$ octupole ($f$) boson. 
The $sdf$-IBM Hamiltonian is then determined by mapping the 
microscopic quadrupole-octupole deformation energy surface onto the
expectation value of the Hamiltonian in the $sdf$ boson condensate
state. 
The boson-fermion coupling Hamiltonian that includes both quadrupole and
octupole boson degrees of freedom contains strength parameters adjusted 
to reproduce low-energy states in the odd-mass nucleus.

The present study is focused on spectroscopic properties of
neutron-rich odd-mass nuclei $^{143,145,147}$Ba. 
For the nucleus $^{145}$Ba, in particular, recent experiments suggest
that there are no signatures of static octupole deformation in the ground- and
low-lying states \cite{zhu1999,rzacaurban2012}, 
even though the neighbouring even-even nucleus $^{144}$Ba has long been 
considered as an example of pronounced octupole correlations \cite{leander1985}. 
Spectroscopic data are available in the neighbouring even-even, as well as odd-even Ba isotopes, 
and this experimental information allows us to constrain the
strength parameters for the boson-fermion interaction in an unambiguous
manner, even though the $sdf$-IBFM Hamiltonian contains  
more parameters than the simpler $sd$-IBFM.

In Ref.~\cite{nomura2013oct} the $sdf$-IBM framework, with the
Hamiltonian determined from a $\beta_2\beta_3$-constrained SCMF
calculation, has already been employed in a systematic study of
quadrupole-octupole shape phase transitions in medium-heavy and
heavy even-even nuclei \cite{nomura2014}. 
Therefore, here we can utilize the mapped
$sdf$-IBM Hamiltonian used in \cite{nomura2014} for the description of
the even-even $sdf$-boson core. 
It should be emphasized that IBFM calculations with an octupole-deformed boson core
have rarely been pursued in the literature. 
To the best of our knowledge, it is only in the phenomenological studies of 
Refs.~\cite{chuu1993,alonso1995,singh1998} that octupole bosons were 
explicitly included in the description of interacting boson-fermion systems.

In Sec.~\ref{sec:model} we outline the method to analyze odd-mass systems
with octupole degrees of freedom and briefly discuss the 
strength parameters of the particle-core coupling. 
Results for both the SCMF and the mapped $sdf$-IBM calculations for the
even-even nuclei $^{142,144,146}$Ba are discussed in 
Sec.~\ref{sec:even}. 
In Sec.~\ref{sec:odd} we present results for the odd-mass nuclei
$^{143,145,147}$Ba, including the systematics of low-energy
positive- and negative-parity yrast states and detailed low-energy level
schemes, as well as $E2$ and $E3$ transition rates in
each odd-mass nucleus. 
A particular emphasis is put on the effect of octupole deformation in the
low-lying states of odd-mass Ba nuclei. 
Finally, a short summary and outlook for future studies are included in
Sec.~\ref{sec:summary}.

\section{The model\label{sec:model}}

The starting point of the present analysis is to perform, for each
even-even Ba nucleus, a self-consistent
mean-field (SCMF) axially-symmetric ($\beta_2,\beta_3$) calculation with constraints on the 
mass quadrupole $Q_{20}$ and octupole
$Q_{30}$ moments. The dimensionless shape variables $\beta_\lambda$
($\lambda=2,3$) are associated with the multipole moments
$Q_{\lambda 0}$: 
\begin{eqnarray}
 \beta_\lambda = \frac{4\pi}{3AR^{\lambda}}Q_{\lambda 0}
\end{eqnarray}
with $R=1.2A^{1/3}$ fm. The relativistic Hartree-Bogoliubov model
\cite{vretenar2005} is used to calculate the ($\beta_2, \beta_3$) deformation
energy surface, single-particle energies and particle occupation
numbers, using the
DD-PC1 functional \cite{DDPC1} in the
particle-hole channel, and a 
separable pairing force of finite range \cite{tian2009} in the particle-particle channel. 
These quantities are subsequently used as microscopic input for the phenomenological IBFM Hamiltonian.

The IBFM Hamiltonian that describes the odd-mass system is composed of
the boson-core Hamiltonian $\hat H_B$, the fermion Hamiltonian 
$\hat H_F$, and the Hamiltonian $\hat H_{BF}$ that couples the boson and fermion
degrees of freedom: 
\begin{eqnarray}
 \hat H_{\rm IBFM} = \hat H_B + \hat H_F + \hat H_{BF}
\end{eqnarray}
The $sdf$-IBM Hamiltonian $\hat H_B$ of the quadrupole- as
well as octupole-deformed even-even boson core nucleus reads:
\begin{eqnarray}
\label{eq:ham-sdf}
 \hat H_B=\epsilon_{d}\hat n_{d}+\epsilon_{f}\hat n_{f}+\kappa_{2}\hat
  Q\cdot\hat Q+\alpha\hat L_d\cdot\hat L_d+\kappa_{3}:\hat V_3^{\+}\cdot\hat V_3:
\end{eqnarray}
This form of the boson Hamiltonian has already been employed in
Ref.~\cite{nomura2014}, and it can be derived by projecting a fully-symmetric 
state in the proton-neutron 
$sdf$ IBM-2 space onto the corresponding IBM-1 state \cite{barfield1988}. 
The first and second terms in Eq.~(\ref{eq:ham-sdf}) are the $d$
and $f$ boson number operators, while the third term in the same equation is the
quadrupole-quadrupole interaction with the quadrupole 
operator 
\begin{eqnarray}
\label{eq:quad}
 \hat Q=s^{\dagger}\tilde d+d^{\dagger}\tilde s+\chi_{dd}[d^{\dagger}\times\tilde d]^{(2)}+\chi_{ff}[f^{\dagger}\times\tilde f]^{(2)}. 
\end{eqnarray}
The fourth term in Eq.~(\ref{eq:ham-sdf}) denotes the rotational term
with the angular momentum operator $\hat L_d=\sqrt{10}[d^{\dagger}\times \tilde
  d]^{(1)}$
and, finally, the last term is the octupole-octupole interaction written
in the normal-ordered form with $\hat V^\+_3$ given by 
\begin{eqnarray}
 \hat V_3^{\+}=s^{\dagger}\tilde
  f+\chi_{df}[d^{\dagger}\times\tilde
  f]^{(3)}. 
\end{eqnarray}
The parameters of the $sdf$ IBM Hamiltonian ($\epsilon_d$, $\epsilon_f$,
$\kappa_2$, $\chi_{dd}$, $\chi_{ff}$, $\kappa_3$ and $\chi_{df}$) are
obtained, for each considered nucleus, by equating the 
SCMF ($\beta_2, \beta_3$) deformation energy surface to the expectation
value of the $sdf$ IBM Hamiltonian of Eq.~(\ref{eq:ham-sdf}) in the
$sdf$-boson coherent state \cite{ginocchio1980}. Since the $\hat L\cdot\hat L$ 
term does not contribute to the energy
surface, the parameter $\alpha$ is determined separately in such a way 
 that the cranking moment of inertia obtained in the boson coherent
state \cite{schaaser1986} at the equilibrium minimum, is equated to the corresponding
Inglis-Belyaev moment of inertia obtained from the SCMF calculation \cite{nomura2011rot}. 
Here the latter is increased by 30\%, taking into account the well known fact that the
Inglis-Belyaev formula underestimates the empirical moments of inertia. 
The $sdf$ IBM parameters used in this study are listed in Table \ref{tab:paraB}. 
Almost the same values of the boson-core parameters are chosen as those
in Ref.~\cite{nomura2014}, except for the strength parameter $\alpha$. 
For a more detailed account of the mapping procedure in the $sdf$ IBM
framework, the reader is referred to Refs.~\cite{nomura2013oct,nomura2014}.

\begin{table}[hb!]
\caption{\label{tab:paraB} The parameters of the $sdf$ IBM
 Hamiltonian. $\epsilon_d$, $\epsilon_f$, $\kappa_2$,
 $\kappa_3$ and $\alpha$ are in units of MeV, and the others are dimensionless.}
\begin{center}
\begin{tabular}{ccccccccc}
\hline\hline
\textrm{} &
\textrm{$\epsilon_d$} &
\textrm{$\epsilon_f$}&
\textrm{$\kappa_2$} &
\textrm{$\chi_{dd}$}&
\textrm{$\chi_{ff}$}&
\textrm{$\alpha$} &
\textrm{$\kappa_3$}&
\textrm{$\chi_{df}$}\\
\hline
$^{142}$Ba & 0.412 & 0.958 & -0.100 & -1.2 & -1.90 & -0.0030 & 0.030 &
				 -0.8 \\
$^{144}$Ba & 0.433 & 0.710 & -0.098 & -1.3 & -2.70 & -0.0199 & 0.048 &
				 -1.5 \\
$^{146}$Ba & 0.202 & 0.729 & -0.098 & -1.2 & -2.75 & -0.0105 & 0.045 &
				 -1.5 \\
\hline\hline
\end{tabular}
\end{center}
\end{table}

Since here only states with one unpaired fermion are considered 
for the description of the low-energy structure of the odd-even system, 
the single-particle Hamiltonian is simply given by 
$\hat H_F=\sum_j\epsilon_j[a^\+_j\tilde a_j]^{(0)}$, with $\epsilon_j$ the 
spherical single-particle energy for the orbital $j$. The energies are determined by 
the SCMF calculation constrained to zero deformation \cite{nomura2016odd} and, 
together with the corresponding occupation probabilities, are listed in Table~\ref{tab:spe-vv}. 
The fermion valence space for the nuclei $^{143,145,147}$Ba includes all single-particle
levels in the neutron $N=82-126$ major shell, that is, $3p_{1/2}$,
$3p_{3/2}$, $1f_{5/2}$, $1f_{7/2}$, $1h_{9/2}$ and $1i_{13/2}$. 

\begin{table}[htb]
\caption{\label{tab:spe-vv}%
Single-particle energies $\epsilon_j$ and occupation probabilities
 $v^2_j$ of the odd neutron 
 for the odd-mass isotopes $^{143,145,147}$Ba.
}
\begin{center}
\begin{tabular}{p{1.0cm}p{1.0cm}p{1.0cm}p{1.0cm}p{1.0cm}p{1.0cm}p{1.0cm}}
\hline\hline
\multirow{2}{*}{} & \multicolumn{2}{l}{$^{143}$Ba} &
 \multicolumn{2}{l}{$^{145}$Ba} & \multicolumn{2}{l}{$^{147}$Ba} \\
\cline{2-3} 
\cline{4-5}
\cline{6-7}
 & $\epsilon_j$ & $v^2_j$  & $\epsilon_j$ & $v^2_j$ &
 $\epsilon_j$ & $v^2_j$\\
\hline
$3p_{1/2}$ & 3.374 & 0.012 & 3.421 & 0.018 & 3.461 & 0.023 \\
$3p_{3/2}$ & 2.732 & 0.017 & 2.797 & 0.024 & 2.856 & 0.033 \\
$2f_{5/2}$ & 2.464 & 0.032 & 2.537 & 0.045 & 2.605 & 0.059 \\
$2f_{7/2}$ & 0.443 & 0.173 & 0.552 & 0.243 & 0.652 & 0.319 \\
$1h_{9/2}$ & 0.000 & 0.319 & 0.000 & 0.441 & 0.000 & 0.556 \\
$1i_{13/2}$ & 3.333 & 0.020 & 3.378 & 0.028 & 3.417 & 0.036 \\
\hline\hline
\end{tabular}
\end{center}
\end{table}

The boson-fermion interaction term $\hat H_{BF}$ consists of terms that 
represent the coupling of the odd neutron to the $sd$-boson space $\hat H_{BF}^{sd}$,
to the $f$ boson space $\hat H_{BF}^{f}$,
and to the combined $sdf$-boson space $\hat H_{BF}^{sdf}$:
\begin{eqnarray}
\label{eq:bf}
 \hat H_{BF}=\hat H_{BF}^{sd} + \hat H_{BF}^{f} + \hat H_{BF}^{sdf}. 
\end{eqnarray}
The first term in Eq.~(\ref{eq:bf}) reads: 
\begin{eqnarray}
 \label{eq:hbf-sd}
 \hat H_{BF}^{sd}
=&&\sum_{j_aj_b}\Gamma_{j_aj_b}^{sd}\hat
  Q_{sd}\cdot[a^{\dagger}_{j_a}\times\tilde a_{j_b}]^{(2)}
\nonumber \\
&&+\sum_{j_aj_bj_c}\Lambda_{j_aj_bj_c}^{dd}
:[[a_{j_a}^\+\times\tilde d]^{(j_c)} 
\times 
[d^\+\times\tilde a_{j_b}]^{(j_c)}]^{(0)}:
\nonumber \\
&&
+\sum_{j_a} A_{j_a}^d[a^{\+}_{j_a}\times\tilde a_{j_a}]^{(0)}\hat n_d, 
\end{eqnarray}
where the first, second and third terms are referred to as (quadrupole) 
dynamical, exchange and monopole terms, respectively \cite{scholten1985,IBFM}. 
$\hat Q_{sd}$ is the $sd$ part of the quadrupole operator in Eq.~(\ref{eq:quad}). 
In the following, single-particle orbitals are denoted by $j_a, j_b,
j_c, \ldots$, while primed ones, such as $j_a', j_b', j_c' \ldots$,
stand for those with opposite parity, unless otherwise specified. 
In a similar fashion,  the following Hamiltonian
is employed for the $f$-boson part:
\begin{eqnarray}
 \hat H_{BF}^f 
&&= \sum_{j_aj_b}\Gamma^{ff}_{j_aj_b}\hat
  Q_{ff}\cdot[a^{\dagger}_{j_a}\times\tilde a_{j_b}]^{(2)}
\nonumber \\
&&+
\sum_{{j_a}{j_b}{j_c^{\prime}}}
\Lambda_{j_aj_bj_c^{\prime}}^{ff}
:[[a^\+_{j_a}\times\tilde f]^{(j_c^{\prime})}
\times
[f^\+\times \tilde a_{j_b}]^{(j_c^{\prime})}]^{(0)}:
\nonumber \\
&&
+\sum_{{j_a}}A_{j_a}^{f}[a^+_{j_a}\times\tilde a_{j_a}]^{(0)}\hat n_f,
\end{eqnarray}
where $\hat Q_{ff}=\chi_{ff}[f^\+\times\tilde f]^{(2)}$, that is,
the third term of the quadrupole operator in Eq.~(\ref{eq:quad}). 
Finally, $\hat H_{BF}^{sdf}$ in Eq.~(\ref{eq:bf}) reads:  
\begin{eqnarray}
\label{eq:hbf-sdf}
\hat H_{BF}^{sdf} 
&&
= 
\sum_{{j_a}{j_b^{\prime}}}\Gamma_{{j_a}{j_b^{\prime}}}^{sf}
\hat V_3^\+\cdot [a_{j_a}^{\dagger}\times\tilde a_{j_b^{\prime}}]^{(3)}
\nonumber \\
&&+
\sum_{{j_a}{j_b^{\prime}j_c}}
\Lambda_{j_aj_b^{\prime}j_c}^{df}
:[[a_{j_a}^\+\times\tilde d]^{(j_c)}
\times
[f^\+\times \tilde a_{j_b^{\prime}}]^{(j_c)}]^{(0)}:
\nonumber \\
&&
+ (H.C.), 
\end{eqnarray}
where the first term denotes the dynamical octupole term.

It has been shown in Ref.~\cite{scholten1985} that, within the
generalized seniority scheme, simple expressions in terms of 
occupation probabilities of the unpaired fermion can be 
derived for the coefficients of each term in $\hat H^{sd}_{BF}$ in
Eq.~(\ref{eq:hbf-sd}): 
\begin{eqnarray}
\label{eq:coeff_sd}
 &&A_{j}^{d}=-A_0^{d}\sqrt{2j+1} \nonumber \\
 &&\Gamma_{j_aj_b}^{sd}
=\Gamma_0^{sd}\gamma_{j_aj_b}^{(2)}
  \nonumber \\
 &&\Lambda_{j_aj_bj_c}^{dd}
=-2\Lambda_0^{dd}\sqrt{\frac{5}{2j_c+1}}
\beta_{j_aj_c}^{(2)}\beta_{j_bj_c}^{(2)}. 
\end{eqnarray}
Here we include the $f$-boson degree of freedom, and obtain similar expressions
for the boson-fermion coupling constants in $\hat
H_{BF}^f$ and $\hat H_{BF}^{sdf}$. 
For the $f$-boson part:
\begin{eqnarray}
\label{eq:coeff_ff}
&&A_{j}^{f}=-A_0^{f}\sqrt{2j+1} \nonumber \\
&&\Gamma_{j_aj_b}^{ff}
=\Gamma_0^{ff}\gamma_{j_aj_b}^{(2)}
 \nonumber \\
&& \Lambda_{j_aj_bj_c'}^{ff}
=-2\Lambda_0^{ff}\sqrt{\frac{7}{2j_c^{\prime}+1}}
\beta_{j_aj_c'}^{(3)}\beta_{j_bj_c'}^{(3)}
\end{eqnarray}
and for the $sdf$-boson terms:
\begin{eqnarray}
\label{eq:coeff_sf}
 &&\Gamma_{j_aj_b^{\prime}}^{sf}
=\Gamma_0^{sf}\gamma_{j_aj_b'}^{(3)}
  \nonumber \\
 &&\Lambda_{j_aj_b^{\prime}j_c}^{df}
=-2\Lambda_0^{df}\sqrt{\frac{7}{2j_c+1}}
\beta_{j_aj_c}^{(2)}\beta_{j_b'j_c}^{(3)}. 
\end{eqnarray}
Note that $\gamma^{(\lambda)}_{ij}=(u_iu_{j}-v_iv_{j})q_{ij}^{(\lambda)}$ 
and
$\beta_{ij}^{(\lambda)}=(u_iv_{j}+u_{j}v_i)q_{ij}^{(\lambda)}$,
where $q_{ij}^{(\lambda)}$ represents the matrix
element of fermion quadrupole ($\lambda=2$) or octupole ($\lambda=3$)
operators in the single-particle basis.  

By following the procedure of Ref.~\cite{nomura2016odd}, the occupation
probabilities $v^2_j$ of the odd particle in the spherical orbital $j$, which appear in
Eqs.~(\ref{eq:coeff_sd})--(\ref{eq:coeff_sf}), are determined by the SCMF 
calculation constrained to zero deformation. 
The $v_j^2$ values used in the present study for the odd-mass nuclei
$^{143,145,147}$Ba are listed in Table~\ref{tab:spe-vv}.

There are altogether fourteen strength parameters for the boson-fermion
interaction that have to be adjusted to the
spectroscopic data for the odd-mass Ba isotopes: six ($\Gamma_0^{sd}$,
$\Gamma_0^{ff}$, $\Lambda_0^{dd}$, 
$\Lambda_0^{ff}$, $A_0^d$ and $A_0^f$) for each of the 
normal-parity $3p_{1/2,3/2}2f_{5/2,7/2}1h_{9/2}$ and the unique-parity
$1i_{13/2}$ single-particle configurations, and two additional parameters $\Gamma_0^{sf}$ and $\Lambda_0^{df}$.
Among the considered odd-mass Ba isotopes, more experimental information about
low-lying states is available for $^{145}$Ba \cite{rzacaurban2012}. 
Thus our strategy is to first
determine the parameters to reproduce several key features of low-energy spectroscopic
data for $^{145}$Ba. Then, under the assumption that the parameters change
gradually with nucleon number, we determine the parameters for the
neighbouring isotopes $^{143}$Ba and $^{147}$Ba. 
Experiments have suggested that the ground state of $^{145}$Ba is
mostly characterized by quadrupole deformation, and there is no strong
coupling with the octupole shape variable \cite{rzacaurban2012}. 
A similar scenario has been suggested for $^{143,147}$Ba \cite{rzacaurban2012} and, 
therefore, we determine the strength parameters for the quadrupole and
octupole modes separately. 
The fitting protocol for $^{145}$Ba is as follows: 
\begin{enumerate}

 \item The strength parameters for the quadrupole part $\hat
       H_{BF}^{sd}$ ($\Gamma_0^{sd}$, 
       $\Lambda_0^{dd}$ and $A_0^d$) are determined to reproduce (i)
       the excitation spectra of 
       the lowest band for each parity as well as (ii) the energy difference
       between the lowest states of each parity: $E({5/2}^-_1)$ for
       negative parity,
       and $E({9/2}^+_1)$ for positive parity.  

 \item The strength parameters for the $f$-boson part $\hat H_{BF}^{f}$
       ($\Gamma_0^{ff}$, $\Lambda_0^{ff}$ and $A_0^{f}$) are specifically relevant
       to those states that are built on $f$-boson configurations. They 
       are determined to reproduce the excitation spectra of 
       (i) the ${11/2}^+$ level at 670 keV (bandhead of the second-lowest positive-parity
       band) for the normal-parity $pfh$ configuration, as well as (ii)
       the ${15/2}^-$ level at 1226 keV for 
       the unique-parity $1i_{13/2}$ configuration, both suggested as candidates for 
       octupole states in Ref.~\cite{rzacaurban2012}. 

\item The term $\hat H_{BF}^{sdf}$ is expected to play a minor role in
      low-lying states of the considered odd-mass Ba nuclei
      and, consequently, their 
      strength parameters $\Gamma_0^{sf}$ and $\Lambda_0^{df}$ are included only 
      perturbatively. In the present study, a constant value is chosen for the
      former, while the latter is neglected as it makes little
      contribution to low-energy excitation spectra. 

\end{enumerate}
The adjusted strength parameters for the $^{143,145,147}$Ba nuclei are listed 
in Table~\ref{tab:paraBF}. Most of the parameters exhibit only a
gradual variation with nucleon number. 
The corresponding $sdf$-IBFM Hamiltonian has been numerically diagonalized by
using the computer code ARBMODEL \cite{arbmodel}.

\begin{table}[htb!]
\caption{\label{tab:paraBF} Strength parameters of the boson-fermion
 interaction $\hat H_{BF}$ in Eq.~(\ref{eq:bf}) employed in the present
 calculation for the $^{143,145,147}$Ba nuclei (in MeV
 units). The numbers in the upper (lower) row in each nucleus correspond to 
 the unique-parity (normal-parity) single-particle configurations.}
\begin{center}
\begin{tabular*}{\columnwidth}{p{0.85cm}p{0.85cm}p{0.85cm}p{0.85cm}p{0.85cm}p{0.85cm}p{0.85cm}p{0.85cm}p{0.85cm}}
\hline\hline
\textrm{} &
\textrm{$\Gamma^{sd}_0$} &
\textrm{$\Gamma^{ff}_0$}&
\textrm{$\Lambda^{sd}_0$} &
\textrm{$\Lambda^{ff}_0$}&
\textrm{$A^{d}_0$}&
\textrm{$A^{f}_0$}&
\textrm{$\Gamma^{sf}_0$}&
\textrm{$\Lambda^{df}_0$}\\
\hline
\multirow{2}{*}{$^{143}$Ba} & 1.40 & 1.20 & 1.0 & 0.0 & -0.75 & -0.75 &
			     \multirow{2}{*}{0.75} &
				 \multirow{2}{*}{0.0} \\
& 0.35 & 0.12 & 1.1 & 1.1 & -1.3 & -1.3 & & \\[0.5em]

\multirow{2}{*}{$^{145}$Ba} & 1.40 & 1.20 & 1.0 & 0.0 & -0.80 & 0.0 &
			     \multirow{2}{*}{0.75} &
				 \multirow{2}{*}{0.0} \\
& 0.40 & 0.13 & 1.0 & 0.30 & -1.0 & -0.15 & & \\[0.5em]

\multirow{2}{*}{$^{147}$Ba} & 1.40 & 1.20 & 1.0 & 0.0 & -0.85 & 0.0 &
			     \multirow{2}{*}{0.75} &
				 \multirow{2}{*}{0.0} \\
& 0.45 & 0.15 & 0.60 & 0.60 & -1.0 & -0.30 & & \\
\hline\hline
\end{tabular*}
\end{center}
\end{table}

Electromagnetic transition probabilities analyzed 
in the present work are the electric quadrupole (E2) and octupole (E3). 
The $E\lambda$ ($\lambda=2,3$) operator is composed of both the boson and
fermion contributions: 
\begin{eqnarray}
 \hat T^{(E\lambda)} = \hat T^{(E\lambda)}_B + \hat T^{(E\lambda)}_F.
\end{eqnarray}
For the E2 operator, the bosonic part reads $\hat
T^{(E2)}_B=e_B^{(2)}\hat Q$, with the quadrupole operator  
$\hat Q$ defined in Eq.~(\ref{eq:quad}), and the fermion E2
operator 
\begin{eqnarray}
 \hat T^{(E2)}_F=
-e_F^{(2)}\sum_{j_aj_b}
\frac{1}{\sqrt{5}}
\gamma_{j_a,j_b}^{(2)}
[a^{\+}_{j_a}\times\tilde a_{j_b}]^{(2)}. 
\end{eqnarray}
$e^{(2)}_B$ and $e^{(2)}_F$ denote bosonic and fermion E2 effective
charges, respectively, and the values $e^{(2)}_B=0.1108\,eb$ and
$e^{(2)}_F=0.5\,eb$ are used for all the considered Ba isotopes. 
$e^{(2)}_B$ has been determined to reproduce the 
experimental value \cite{bucher2016} of the $B(E2; 2^+_1\rightarrow 0^+_1)$ transition
rate in the even-even nucleus $^{144}$Ba. 
Similarly, for the E3 transition operator, the bosonic part reads 
$\hat T^{(E3)}=e^{(3)}_B(\hat V_3^\+ + \hat V_3)$, and the fermion part
can be written, in analogy to the quadrupole 
one, as
\begin{eqnarray}
 \hat T^{(E3)}_F = -e_F^{(3)}\sum_{j_aj_b'}
\frac{1}{\sqrt{7}}
\gamma_{j_a,j_b'}^{(3)}
[a^{\+}_{j_a}\times\tilde a_{j_b'}]^{(3)}. 
\end{eqnarray}
The E3 boson effective charge of $e^{(3)}_B=0.09\,eb^{3/2}$ is 
taken from our previous calculation of the same even-even Ba nuclei
\cite{nomura2014}, while the E3 fermion charge of 
$e^{(3)}_F=0.5\,eb^{3/2}$ is employed in the present calculation.

\section{Results for the even-even Ba isotopes\label{sec:even}}

\begin{figure}[htb!]
\begin{center}
\includegraphics[width=\linewidth]{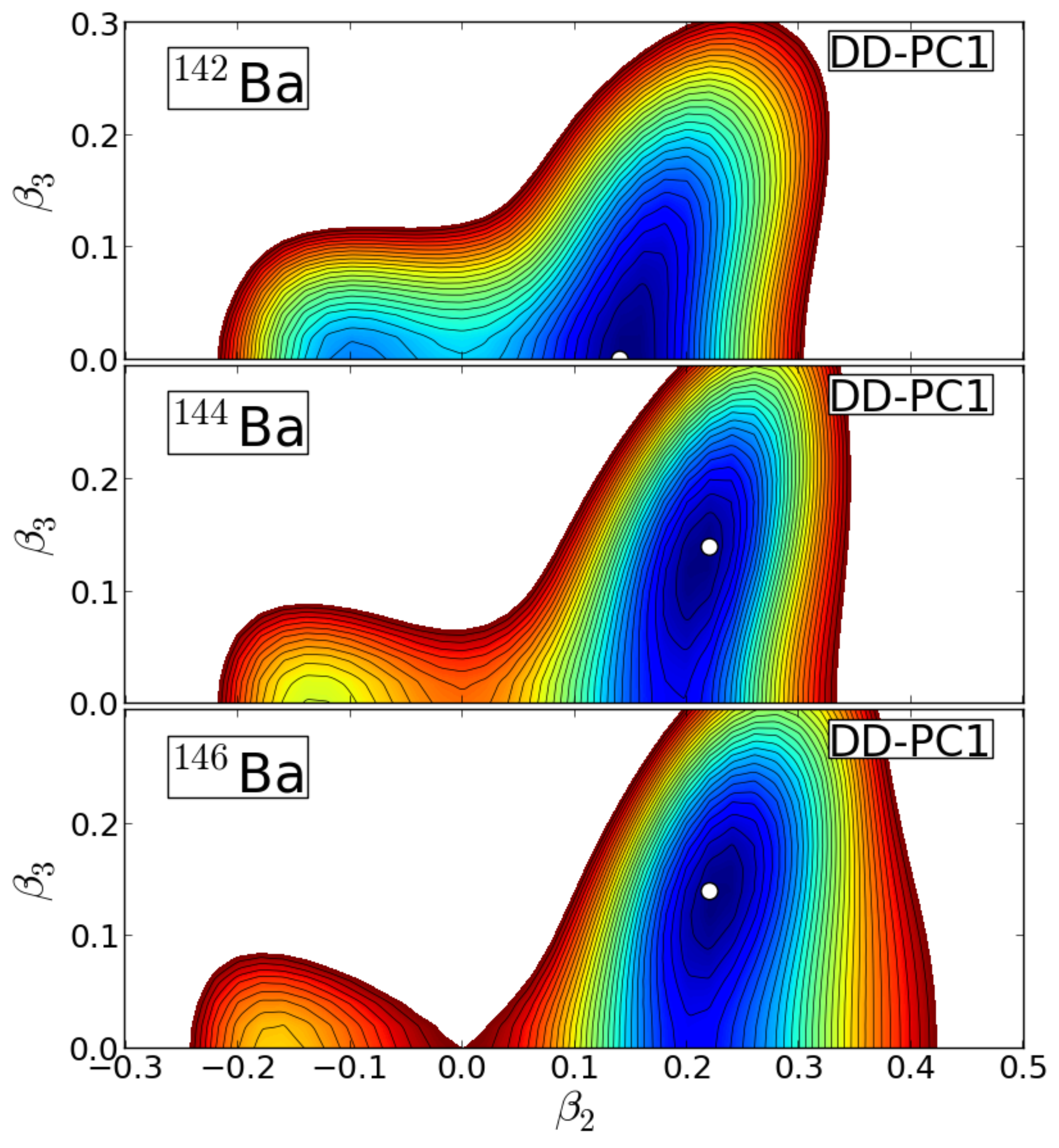}
\caption{(Color online) The SCMF ($\beta_{2}, \beta_{3}$) deformation energy
 surfaces for $^{142,144,146}$Ba, obtained with the DD-PC1 nuclear functional \cite{DDPC1}
and a separable pairing force of finite range \cite{tian2009}. The energy
 difference between neighbouring contours is 200 keV. Equilibrium minima
 are identified by open circles.} 
\label{fig:pes}
\end{center}
\end{figure}

\begin{figure}[htb!]
\begin{center}
\includegraphics[width=\linewidth]{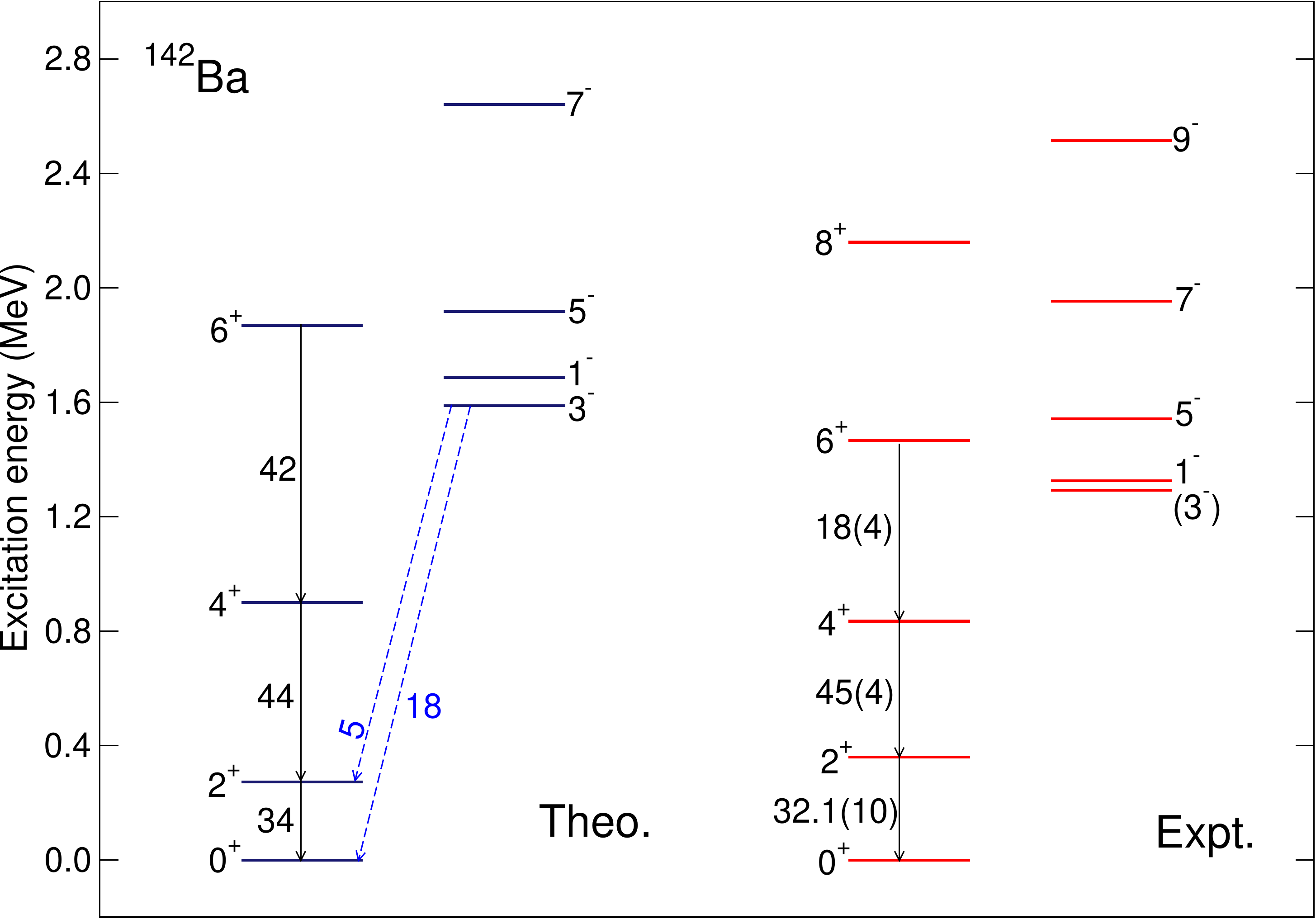}
\includegraphics[width=\linewidth]{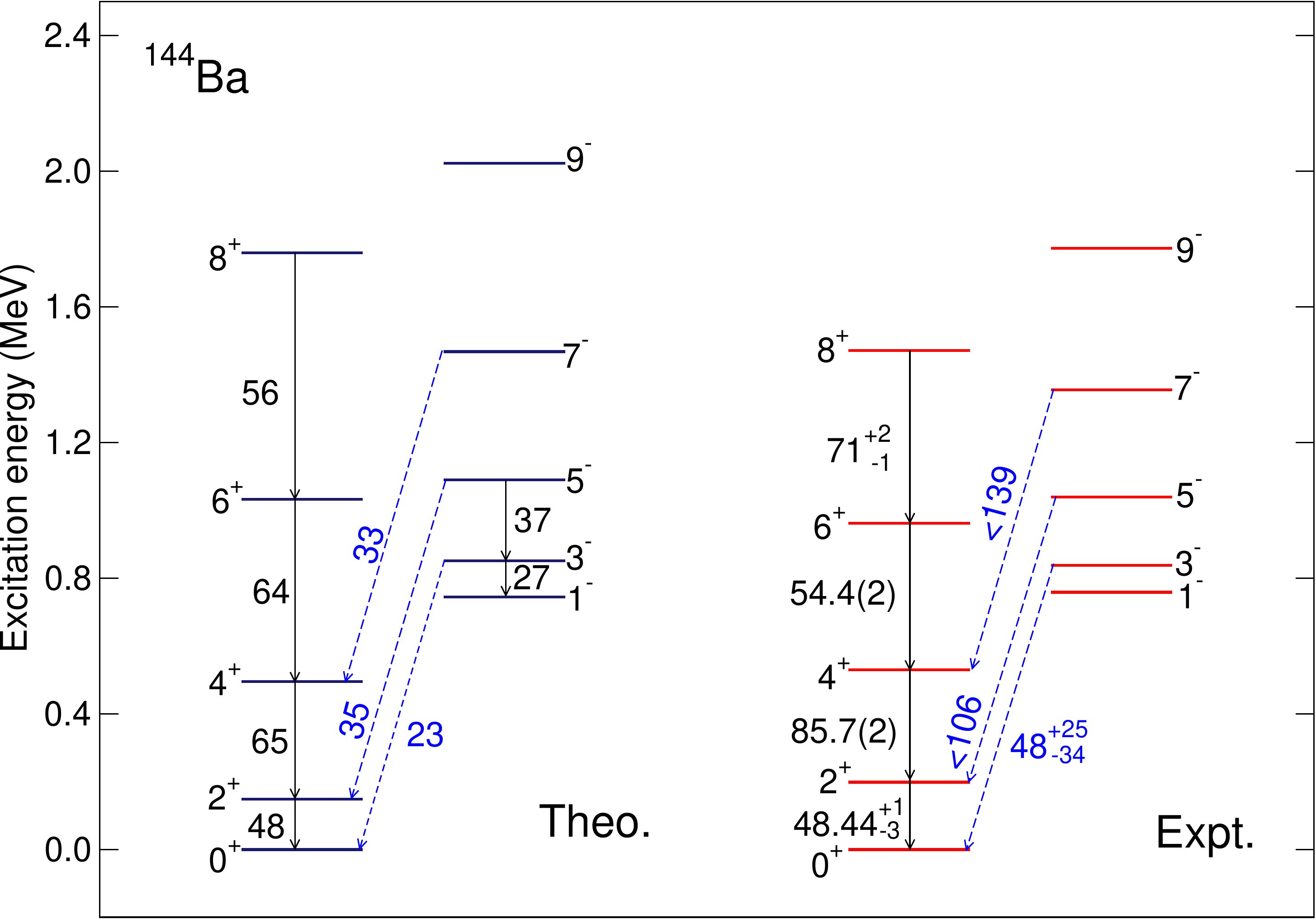}
\includegraphics[width=\linewidth]{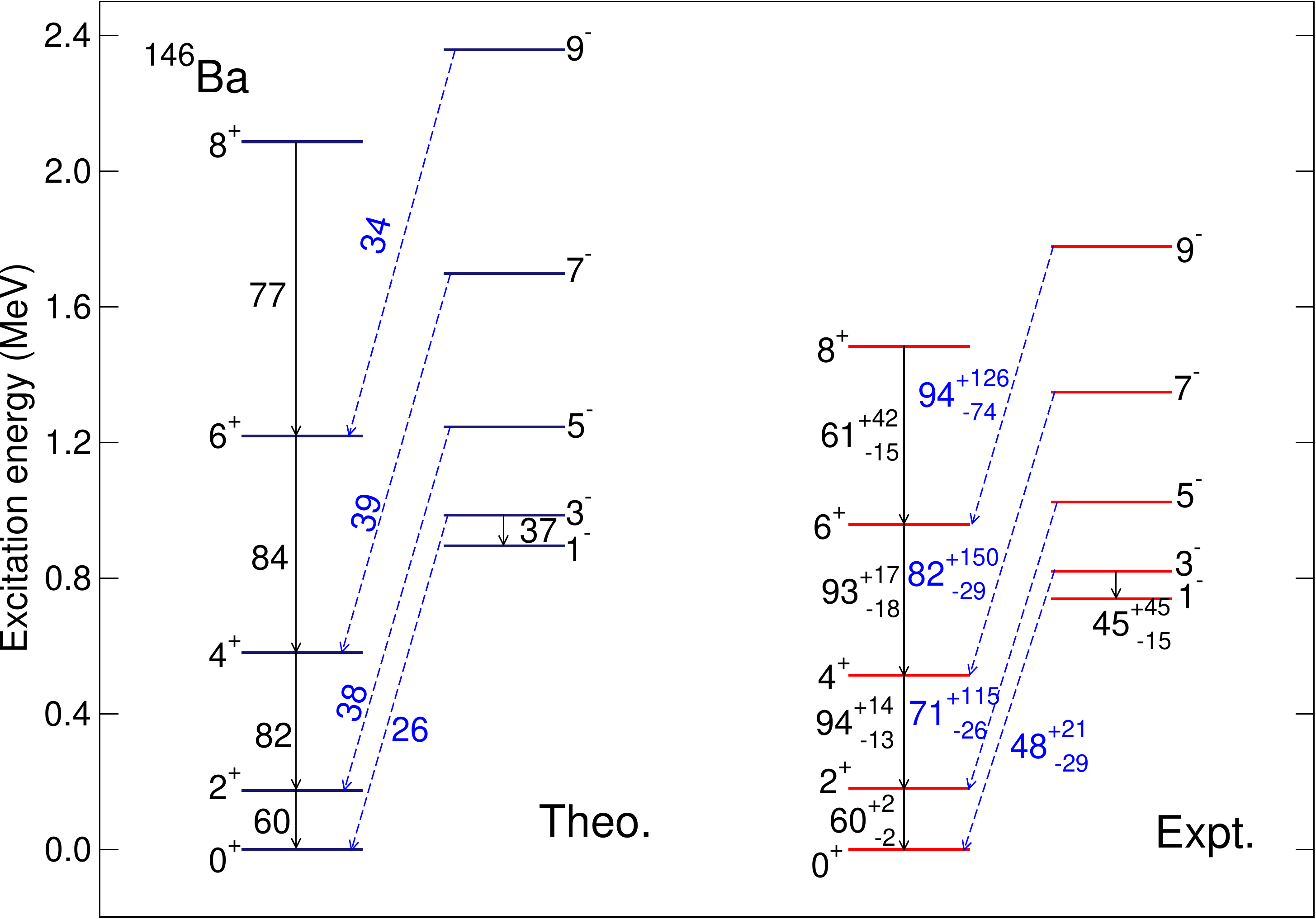}
\caption{(Color online) Low-energy positive- and negative-parity spectra
 of the even-even boson core nuclei $^{142,144,146}$Ba. The $B(E2)$
 (numbers along arrows within each band) and
 $B(E3)$ (inter-band, dashed arrows) values are given  
  in Weisskopf units. Experimental values are taken from
 Ref.~\cite{data} ($^{142}$Ba), Ref.~\cite{bucher2016} 
 ($^{144}$Ba) and Ref.~\cite{bucher2017} ($^{146}$Ba).} 
\label{fig:level-even}
\end{center}
\end{figure}

We first briefly discuss the results obtained for the even-even nuclei
$^{142,144,146}$Ba. 
Figure~\ref{fig:pes} depicts the axially-symmetric
($\beta_{2},\beta_{3}$) deformation energy surfaces calculated 
with the constrained relativistic
Hartree-Bogoliubov method. 
For $^{142}$Ba the equilibrium minimum is found on the $\beta_3=0$ axis,
indicating that it has a weakly-deformed quadrupole shape. 
In $^{144,146}$Ba a minimum with non-zero $\beta_3$ deformation
($\beta\approx 0.1$) appears. The minimum is not very pronounced and is rather soft in
the $\beta_3$ direction, suggesting the occurrence of octupole vibrational states 
in these nuclei.

The excitation spectra and transition rates for 
$^{142,144,146}$Ba are computed by diagonalizing the IBM Hamiltonian 
Eq.~(\ref{eq:ham-sdf}), determined from  
the SCMF ($\beta_{2},\beta_{3}$) deformation energy surface, in the $sdf$-boson basis. 
The low-energy level schemes for the $^{142,144,146}$Ba isotopes are
displayed in Fig.~\ref{fig:level-even}. In general, the theoretical predictions 
are in good agreement with the
experimental results \cite{data,bucher2016,bucher2017}, not only for the
excitation energies but also for the $E2$ and $E3$ transition
strengths. In the transition from  $^{142}$Ba  $^{144}$Ba nucleus, in particular, we note the 
pronounced lowering of the the negative-parity band
(cf.  the corresponding SCMF deformation energy surface in Fig.~\ref{fig:pes}). 
A rather large value $B(E3;3^-\rightarrow 0^+)$ is
predicted for all three even-even Ba nuclei, but still considerably smaller 
than the experimental values reported for $^{144,146}$Ba \cite{bucher2016,bucher2017}. 
Note, however, the large uncertainty of the latter.

\section{Spectroscopic properties of odd-mass Ba isotopes\label{sec:odd}}

\begin{figure}[htb!]
\begin{center}
\includegraphics[width=\linewidth]{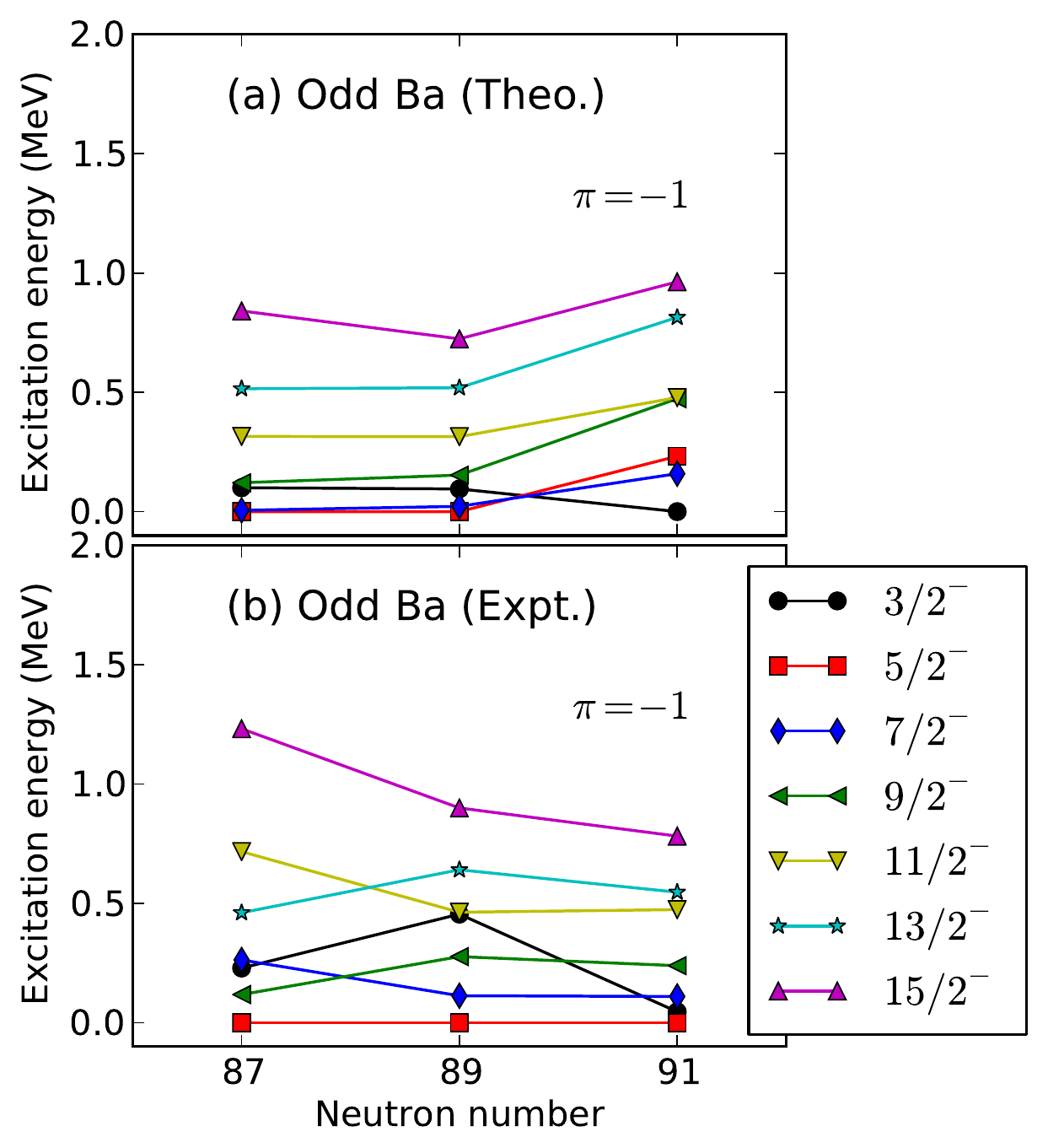}\\
\caption{(Color online) Evolution of low-energy negative-parity levels of the odd-mass Ba isotopes as functions
 of the neutron number. Experimental excitation energies are taken from Refs.~\cite{zhu1999,rzacaurban2012,rzacaurban2013}.} 
\label{fig:level-odd-pfh}
\end{center}
\end{figure}

\begin{figure}[htb!]
\begin{center}
\includegraphics[width=\linewidth]{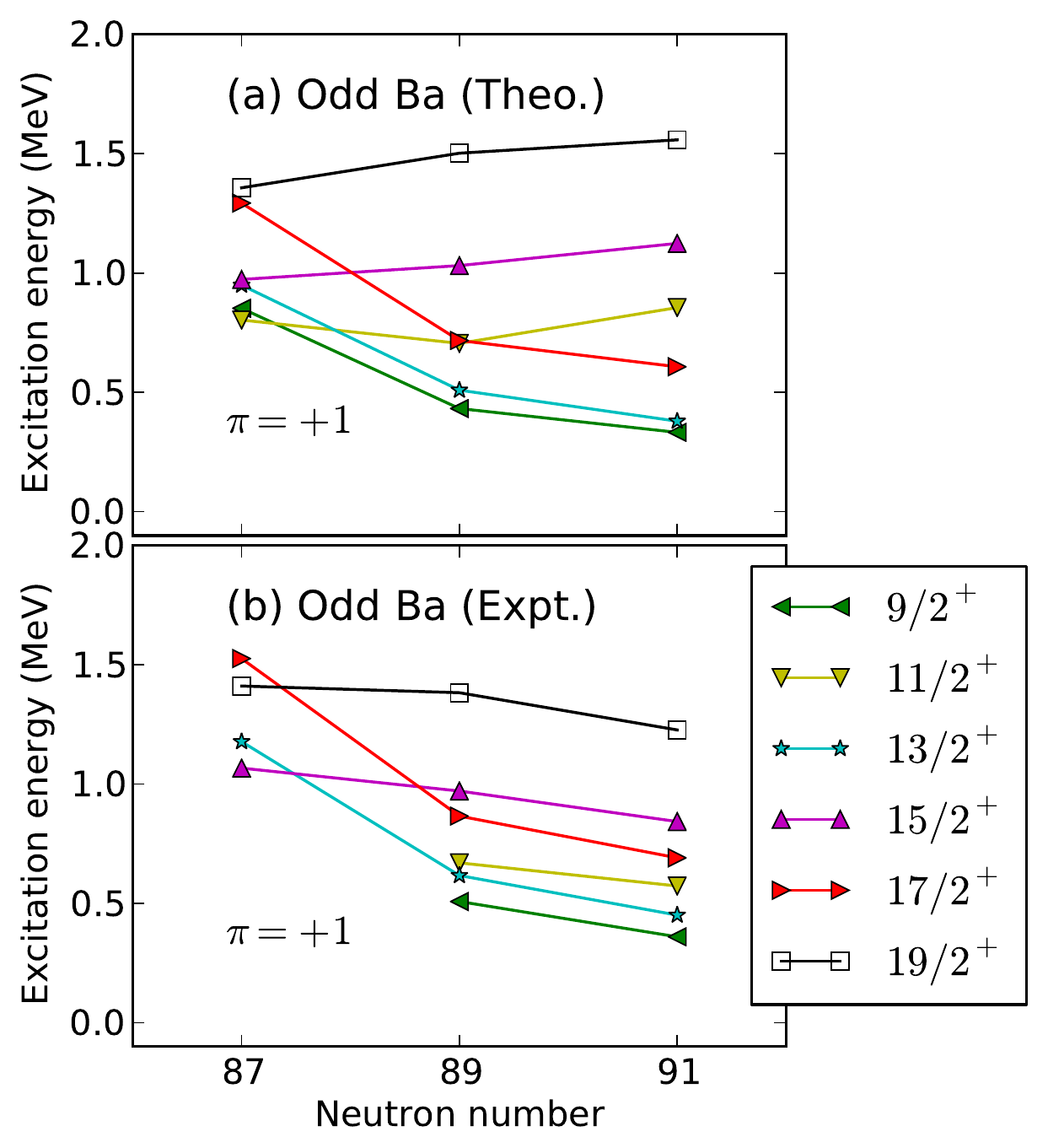}\\
\caption{(Color online) Same as Fig.~\ref{fig:level-odd-pfh}, but for
 the positive-parity levels.} 
\label{fig:level-odd-i13}
\end{center}
\end{figure}

\subsection{Evolution of low-energy excitation spectra}

In Figs.~\ref{fig:level-odd-pfh} and \ref{fig:level-odd-i13} the lowest-energy negative- and
positive-parity states for for each spin of 
$^{143,145,147}$Ba are plotted as functions of the neutron number,
respectively. 
Note that the spin and parity of the lowest ${3/2}^-$ state
for $^{143}$Ba, the ${3/2}^-$, ${11/2}^+$, ${15/2}^+$, ${19/2}^+$ states for
$^{145}$Ba, and the ${3/2}^-$, ${11/2}^-$, ${15/2}^-$, ${11/2}^+$, ${15/2}^+$ and 
${19/2}^+$ states have been assigned tentatively 
\cite{zhu1999,rzacaurban2012,rzacaurban2013,data}. 
The excitation spectra obtained by the diagonalization
of the $sdf$-IBFM Hamiltonian reproduce the trend of the data,
even though very little variation has been allowed for most of the 
strength parameters for the boson-fermion interaction (Table~\ref{tab:paraBF}). 
In $^{147}$Ba, however, experimentally the spin of the ground state is $J^{\pi}={5/2}^-$
\cite{rzacaurban2013}, whereas in the present calculation it is
$J^{\pi}={3/2}^-$. This state could be among the lowest in the data on 
$^{147}$Ba. 
A signature of shape transition is a rather rapid decrease of the ${9/2}^+$,
${13/2}^+$ and ${17/2}^+$ energy levels from $^{143}$Ba to $^{145}$Ba, 
and somewhat less steep from $^{145}$Ba to $^{147}$Ba. The empirical 
trend is nicely reproduced by the calculation, and  
this correlates with the fact that, in the even-even systems, the
non-zero octupole equilibrium deformation $\beta_3$ appears in the SCMF
energy surface for $^{144}$Ba (Fig.~\ref{fig:pes}). 
There are no distinct irregularities in the excitation spectra shown in
Figs.~\ref{fig:level-odd-pfh} and \ref{fig:level-odd-i13}.

\subsection{Detailed comparison of level schemes}

Figures \ref{fig:ba143-detail}--\ref{fig:ba147-detail} display detailed
comparisons between theoretical and experimental 
positive-parity and negative-parity low-lying bands in
the odd-mass nuclei $^{143,145,147}$Ba. 
In organizing the theoretical level schemes, states are
classified into bands according to the dominant E2 decay rates and the
similarity in the composition of their IBFM wave functions. 
The labels in parentheses denote states that are assigned only
tentatively in experiment.
Also, only experimental states that are classified into bands are
plotted in Figs.~\ref{fig:ba143-detail}--\ref{fig:ba147-detail}. 
Some experimental states that do not belong to these bands, for instance the ${3/2}^-_1$ and
${7/2}^-_1$ states for $^{143}$Ba, are not included in
Fig.~\ref{fig:ba143-detail}, even though they are plotted in Fig.~\ref{fig:level-odd-pfh}. 
In Tables~\ref{tab:frac-143ba}--\ref{tab:frac-147ba} we list the expectation values
of the $f$-boson number operator $\langle\hat n_f\rangle$, as well as the
contribution of each single-particle component in the IBFM wave
function of the band-head state of each band. 
The predicted $B(E2)$ and $B(E3)$ values are included in
Tables~\ref{tab:trans-145ba}--\ref{tab:trans-147ba}. Note that presently 
data are not available for these quantities. 


\subsubsection{$^{143}$Ba}

\begin{figure}[htb!]
\begin{center}
\includegraphics[width=\linewidth]{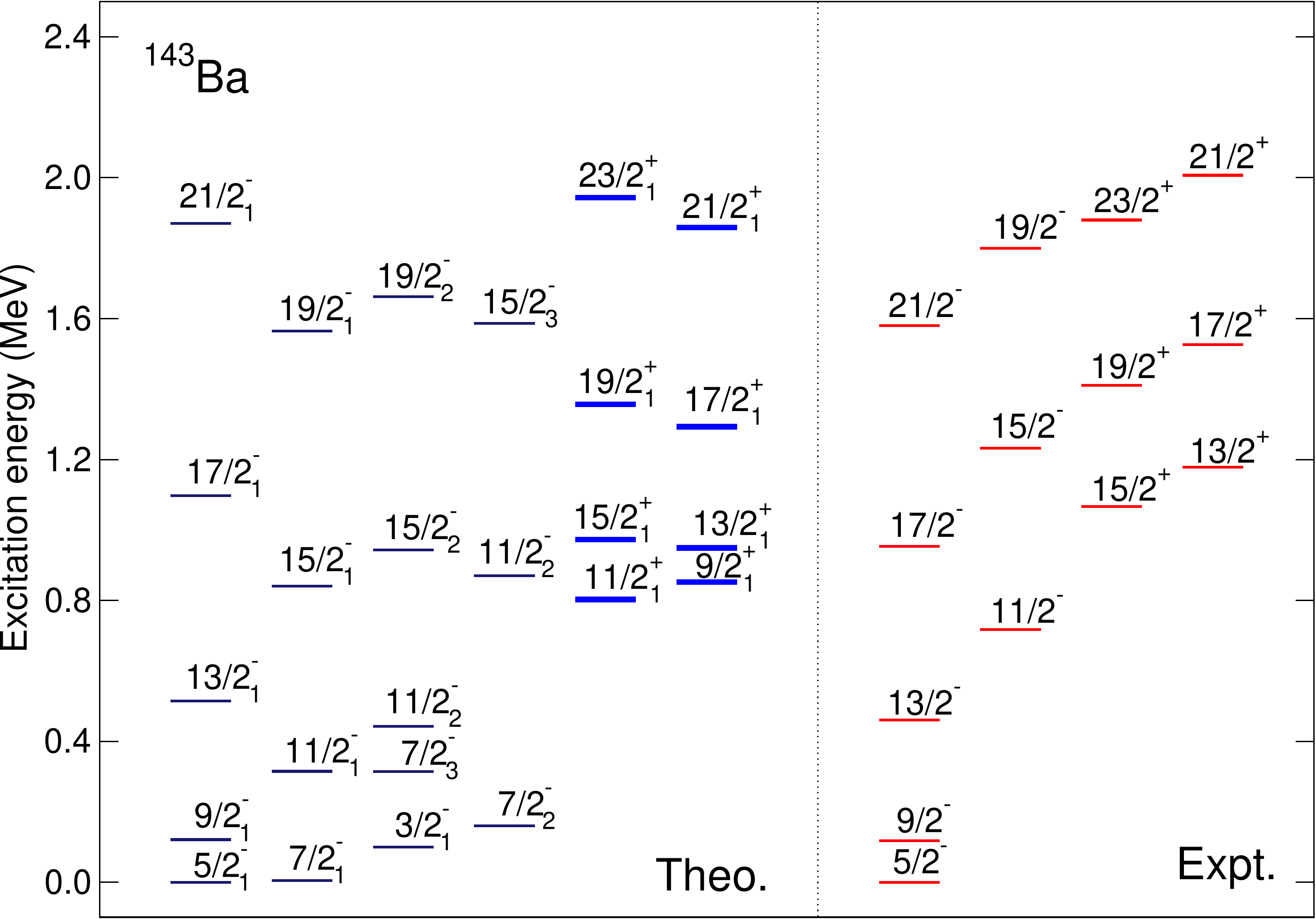}
\caption{(Color online) Detailed comparison of the calculated and
 experimental low-energy positive- and negative-parity spectra
 of $^{143}$Ba. Those theoretical energy levels that are made of one
 $f$-boson configuration are drawn in bold and in color blue. Data are taken from Ref.~\cite{rzacaurban2012}.} 
\label{fig:ba143-detail}
\end{center}
\end{figure}

For the even-even core nucleus $^{142}$Ba the SCMF
($\beta_2, \beta_3$) deformation energy surface exhibits an equilibrium minimum at
axial quadrupole deformation $\beta_2 \approx 0.14$ and octupole deformation  $\beta_3=0$, 
and the corresponding spectra display negative-parity states at relatively high excitation energies 
compared to $^{144,146}$Ba. 
Hence, octupole deformation is expected to play a rather minor role
in the odd-mass system $^{143}$Ba.
Experimentally four bands, two of positive- and two negative-parity each, 
have been established in $^{143}$Ba \cite{zhu1999}. 
The predicted level scheme shown in Fig.~\ref{fig:ba143-detail} reproduces
nicely the lowest negative-parity ground-state band built on the ${5/2}^-_1$
state, as well as the energy of the ${15/2}^+_1$ state, which is the
lowest positive-party state in experiment.

As shown in Table \ref{tab:frac-143ba}, the present calculation does not predict the
presence of octupole states (i.e., states that contain one or more $f$
bosons in their wave functions) in the vicinity of the ground state in $^{143}$Ba: all
the negative-parity bands shown in Fig.~\ref{fig:ba143-detail} are composed 
mainly of the odd neutron in the $pfh$ orbitals coupled to the $sd$-boson
space. On the other hand, the band built on the $11/2^+_1$ state is predicted to
contain predominantly states with one $f$-boson, and similar for the ${9/2}^+_1$ band. 
From Table~\ref{tab:trans-143ba} one notices 
several significant $B(E3)$ transition probabilities from the bands based on the 
$J^{\pi}={9/2}^+_1$ and $11/2^+_1$ states to the low-lying decoupled negative-parity band, e.g.,
$B(E3; {15/2}^+_1\rightarrow {9/2}^-_1) = 11$ W.u. and
$B(E3; {17/2}^+_1\rightarrow {11/2}^-_1) = 13$ W.u.

\begin{table}[htb!]
\caption{\label{tab:trans-143ba} Predicted $B(E2)$ and $B(E3)$ values (in Weisskopf units) 
for transitions among low-energy states of $^{143}$Ba.}
\begin{center}
\begin{tabular}{lc}
\hline\hline
\textrm{$B(E\lambda; J_i^\pi \rightarrow J_f^\pi)$} &
\textrm{Theory (W.u.)} \\
\hline
$B(E2; {9/2}^-_1 \rightarrow {5/2}^-_1)$ & 16 \\
$B(E2; {11/2}^-_1 \rightarrow {7/2}^-_1)$ & 25 \\
$B(E2; {13/2}^-_1 \rightarrow {9/2}^-_1)$ & 35 \\
$B(E2; {13/2}^+_1 \rightarrow {9/2}^+_1)$ & 21 \\
$B(E2; {15/2}^+_1 \rightarrow {11/2}^+_1)$ & 31 \\
$B(E2; {11/2}^+_2 \rightarrow {7/2}^+_1)$ & 4.2 \\
$B(E3; {11/2}^+_1 \rightarrow {5/2}^-_1)$ & 5.5 \\
$B(E3; {15/2}^+_1 \rightarrow {9/2}^-_1)$ & 11 \\
$B(E3; {13/2}^+_1 \rightarrow {7/2}^-_1)$ & 8.4 \\
$B(E3; {17/2}^+_1 \rightarrow {11/2}^-_1)$ & 13 \\
\hline\hline
\end{tabular}
\end{center}
\end{table}

\begin{table}[htb!]
\caption{\label{tab:frac-143ba} Expectation value of the $f$-boson
 number operator $\langle\hat n_f\rangle$, and squares of the amplitudes (in percentage) of each 
 spherical single-particle
 configuration in the IBFM wave functions of band-head
 states in $^{143}$Ba (cf. Fig.~\ref{fig:ba143-detail}.)}
\begin{center}
\begin{tabular}{cccccccc}
\hline\hline
\textrm{$J^{\pi}$} &
\textrm{$\langle\hat n_f\rangle$} &
\textrm{$3p_{1/2}$} &
\textrm{$3p_{3/2}$} &
\textrm{$2f_{5/2}$} &
\textrm{$2f_{7/2}$} &
\textrm{$1h_{9/2}$} &
\textrm{$1i_{13/2}$} \\
\hline
${5/2}^-_1$ & 0.000 & 0 & 0 & 4 & 4 & 92 & 0 \\
${7/2}^-_1$ & 0.000 & 0 & 0 & 2 & 5 & 93 & 0 \\
${3/2}^-_1$ & 0.000 & 0 & 9 & 0 & 78 & 13 & 0 \\
${7/2}^-_2$ & 0.001 & 0 & 9 & 0 & 80 & 11 & 0 \\
${11/2}^+_1$ & 1.000 & 0 & 0 & 2 & 2 & 96 & 0 \\
${9/2}^+_1$ & 1.000 & 0 & 0 & 2 & 3 & 95 & 0 \\
\hline\hline
\end{tabular}
\end{center}
\end{table}


\subsubsection{$^{145}$Ba}

\begin{figure}[htb!]
\begin{center}
\includegraphics[width=\linewidth]{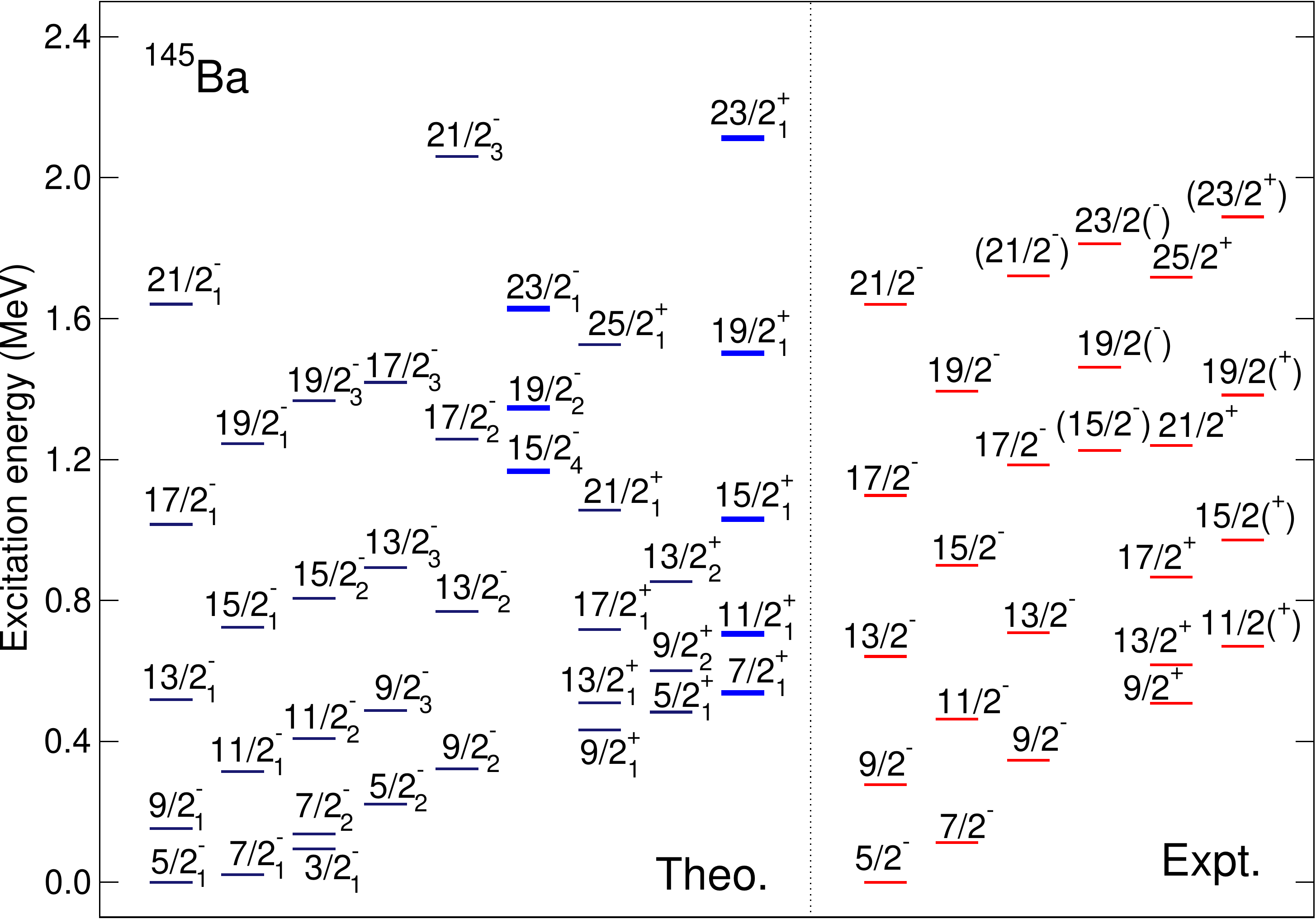}
\caption{(Color online) Same as in the caption to Fig.~\ref{fig:ba143-detail}, but for the nucleus
 $^{145}$Ba. Experimental excitation energies are from Ref.~\cite{rzacaurban2012}.} 
\label{fig:ba145-detail}
\end{center}
\end{figure}

More experimental information is available on the isotope $^{145}$Ba and, 
as discussed in Sec.~\ref{sec:even}, since the corresponding even-even boson core nucleus 
$^{144}$Ba exhibits an octupole-soft potential at the SCMF level (cf. 
Fig.~\ref{fig:pes}), we also expect that octupole correlations play a more important role in the 
low-energy spectra of this nucleus. The calculated excitation spectrum is compared to 
the corresponding experimental bands in Fig.~\ref{fig:ba145-detail}. 
As shown in Table~\ref{tab:frac-145ba}, the lowest two negative-parity bands
in $^{145}$Ba are built on the 
${5/2}^-_1$ and ${7/2}^-_1$ states that are characterized by the coupling of the 
unpaired neutron in the $1h_{9/2}$
single-particle orbital to the $sd$ boson space. The lowest
positive-parity state ${9/2}^+_1$ is described
by the coupling of the $1i_{13/2}$ orbital to $sd$-boson states. 

From Table~\ref{tab:frac-145ba} it follows that the theoretical negative-parity band
built on the ${15/2}^-_4$ state (calculated at 1167 keV) is dominated by 
the coupling of the $1i_{13/2}$ single-particle orbital to states with one
$f$-boson. Theoretically the ${15/2}^-_4$ state appears to be an
octupole state and is located close to the experimental
${15/2}^-$ state found at 1226 keV \cite{rzacaurban2012}. 
Moreover, rather strong E3 transitions from the $J^\pi={15/2}^-_4$ band to the
corresponding low-lying positive-parity bands are predicted: for
instance $B(E3; {15/2^-_4}\rightarrow {9/2^+_1})=25$ and  $B(E3; {19/2^-_2}\rightarrow
{13/2^+_1})=31$ (in W.u), comparable to 
the $B(E3; 3^-_1\rightarrow 0^+_1)$ value of 23 W.u. in the corresponding even-even
core nucleus $^{144}$Ba (see, Fig.~\ref{fig:level-even}). 
However, to verify model predictions, experimental information on 
the $B(E2)$ and $B(E3)$ values is needed.

For positive parity, the theoretical band built on the ${7/2}^+_1$ in
$^{145}$Ba corresponds to the coupling of the $1h_{9/2}$ single-neutron 
configuration to states with one $f$-boson (cf. 
Table~\ref{tab:frac-145ba}). The theoretical 
${11/2}^+_1$ level, calculated at 705 keV, can be compared with the
experimental ${11/2}^+$ state at 670 keV \cite{rzacaurban2012}, which
has been suggested as a candidate for an octupole state. 
Non-negligible E3 transition strength from the ${7/2}^+_1$ band to the
negative-parity ground-state band is predicted in the present
calculation: $B(E3; {11/2^+_1}\rightarrow {5/2^-_1})=4.7$ and $B(E3;
{15/2^+_1}\rightarrow {9/2^-_1})=15$ (in W.u).

\begin{table}[htb!]
\caption{\label{tab:trans-145ba} Same as in the caption to Table~\ref{tab:trans-143ba},
 but for the $^{145}$Ba nucleus.}
\begin{center}
\begin{tabular}{lc}
\hline\hline
\textrm{$B(E\lambda; J_i^{\pi}\rightarrow J_f^{\pi}$)} &
\textrm{Theory (W.u.)} \\
\hline
$B(E2; {7/2}^-_1 \rightarrow {5/2}^-_1)$ & 23 \\
$B(E2; {9/2}^-_1 \rightarrow {5/2}^-_1)$ & 10 \\
$B(E2; {11/2}^-_1 \rightarrow {7/2}^-_1)$ & 23 \\
$B(E2; {13/2}^-_1 \rightarrow {9/2}^-_1)$ & 38 \\
$B(E2; {11/2}^+_1 \rightarrow {7/2}^+_1)$ & 21 \\
$B(E2; {13/2}^+_1 \rightarrow {9/2}^+_1)$ & 75 \\
$B(E2; {15/2}^+_1 \rightarrow {11/2}^+_1)$ & 43 \\
$B(E2; {17/2}^+_1 \rightarrow {13/2}^+_1)$ & 78 \\
$B(E3; {11/2}^+_1 \rightarrow {5/2}^-_1)$ & 4.7 \\
$B(E3; {15/2}^+_1 \rightarrow {9/2}^-_1)$ & 15 \\
$B(E3; {13/2}^+_1 \rightarrow {7/2}^-_1)$ & 0.028 \\
$B(E3; {17/2}^+_1 \rightarrow {11/2}^-_1)$ & 0.59 \\
$B(E3; {15/2}^-_1 \rightarrow {9/2}^+_1)$ & 0.00014 \\
$B(E3; {15/2}^-_2 \rightarrow {9/2}^+_1)$ & 0.00078 \\
$B(E3; {15/2}^-_4 \rightarrow {9/2}^+_1)$ & 25 \\
$B(E3; {19/2}^-_2 \rightarrow {13/2}^+_1)$ & 31 \\
\hline\hline
\end{tabular}
\end{center}
\end{table}

\begin{table}[htb!]
\caption{\label{tab:frac-145ba} Same as in the caption to Table~\ref{tab:frac-143ba}, but
 for $^{145}$Ba.}
\begin{center}
\begin{tabular}{cccccccc}
\hline\hline
\textrm{$J^{\pi}$} &
\textrm{$\langle\hat n_f\rangle$} &
\textrm{$3p_{1/2}$} &
\textrm{$3p_{3/2}$} &
\textrm{$2f_{5/2}$} &
\textrm{$2f_{7/2}$} &
\textrm{$1h_{9/2}$} &
\textrm{$1i_{13/2}$} \\
\hline
${5/2}^-_1$ & 0.000 & 0 & 0 & 5 & 3 & 92 & 0 \\
${7/2}^-_1$ & 0.000 & 0 & 0 & 1 & 4 & 95 & 0 \\
${3/2}^-_1$ & 0.005 & 3 & 17 & 1 & 67 & 11 & 0 \\
${5/2}^-_2$ & 0.002 & 1 & 9 & 1 & 70 & 20 & 0 \\
${9/2}^-_2$ & 0.000 & 0 & 1 & 2 & 8 & 89 & 0 \\
${15/2}^-_4$ & 0.998 & 0 & 0 & 0 & 0 & 0 & 100 \\
${9/2}^+_1$ & 0.000 & 0 &0 & 0 & 0 & 0 & 100 \\
${5/2}^+_1$ & 0.005 & 0 & 0 & 0 & 0 & 0 & 99 \\ 
${7/2}^+_1$ & 1.000 & 0 & 0 & 1 & 4 & 95 & 0 \\
\hline\hline
\end{tabular}
\end{center}
\end{table}


\subsubsection{$^{147}$Ba}

\begin{figure}[htb!]
\begin{center}
\includegraphics[width=\linewidth]{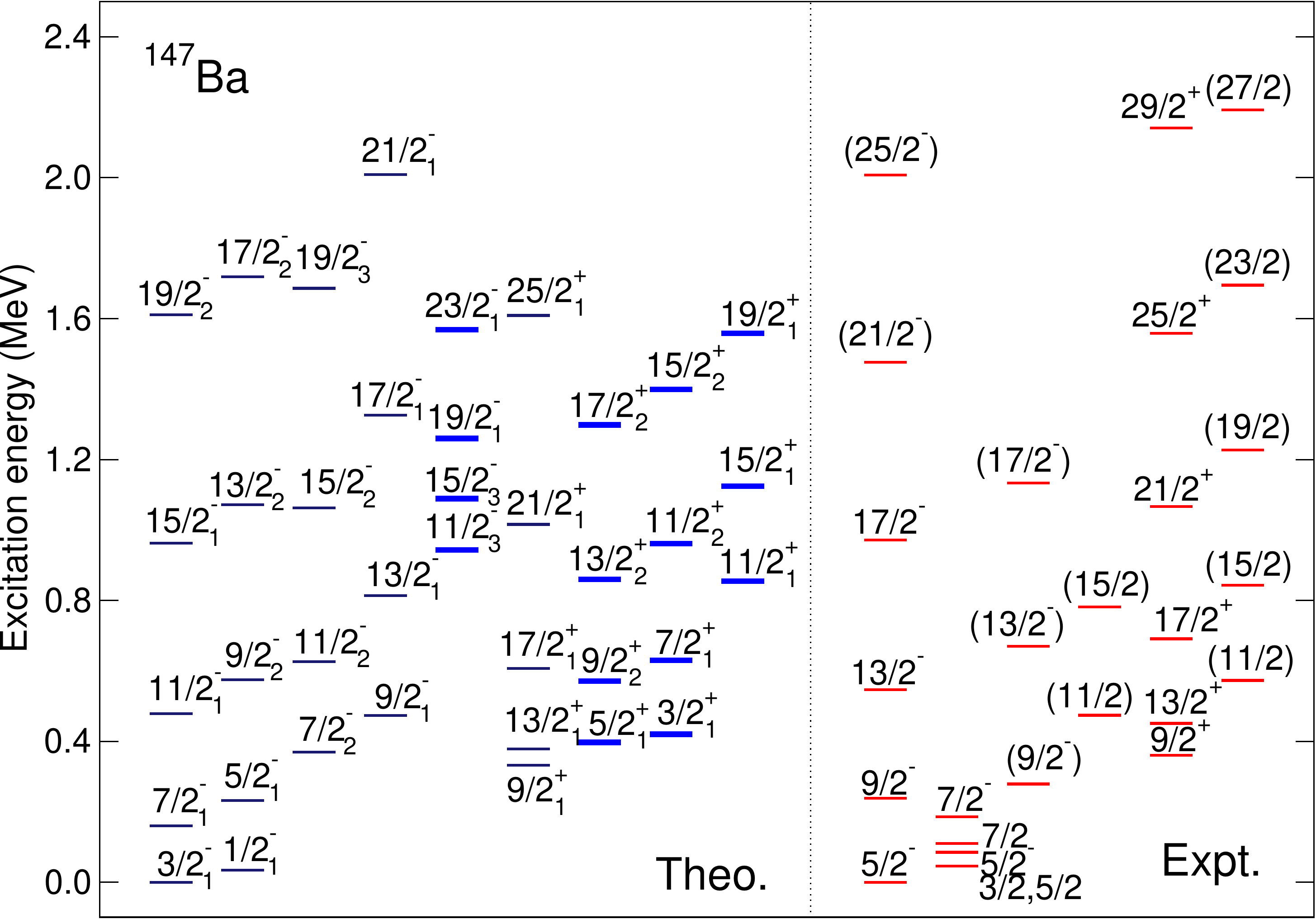}
\caption{(Color online) Same as in the caption to Fig.~\ref{fig:ba143-detail}, but for
 $^{147}$Ba and with data from Ref.~\cite{rzacaurban2013}.} 
\label{fig:ba147-detail}
\end{center}
\end{figure}

From the SCMF ($\beta_2, \beta_3$) deformation energy surfaces of $^{144}$Ba and $^{146}$Ba 
(cf. Fig.~\ref{fig:pes}) one expects   
rather similar low-energy excitation spectra in their odd-$N$ neighbours $^{145}$Ba and 
$^{147}$Ba, respectively. This is, to a certain extent, observed in the experimental spectra 
\cite{rzacaurban2013}, even though the bands of $^{147}$Ba appear more compressed, 
as seen when comparing the data in Figs.~\ref{fig:ba145-detail} and
\ref{fig:ba147-detail}. 
The ground-state spin ${5/2}^-$ has been identified for $^{147}$Ba, just as in the $^{143,145}$Ba 
neighbours \cite{rzacaurban2013}. 
This is, however, at variance within the present calculation, which predicts ${3/2}^-$ for the ground state. 
%
In fact, it is possible to reproduce the ground state spin of
$J={5/2}^-$ by playing with the parameters. In that case, however, one would
have to choose an unrealistic value for the boson-fermion interaction
strength, for instance, negative value for $\Gamma^{(sd)}_0$. 
The wrong sign implies that the single-particle energies
and/or occupation probabilities used in the present calculation are not
necessarily optimal. We then checked that decreasing the occupation probability for
the $1h_{9/2}$ orbital, for instance, by 25 \%, allowed to reproduce the
correct level ordering of the ground state, but such an adjustment is
not justified within the present framework and is beyond the scope of this paper.  
Nevertheless, one notices in Fig.~\ref{fig:ba147-detail} that a low-lying
$J={3/2}$ level could be also present in experiment
\cite{rzacaurban2013}, i.e., among a set of levels close to the 
$J^{\pi}={5/2}^-$ ground state, even though its spin and parity have not
been firmly established. 
Table~\ref{tab:frac-147ba} also shows that the structure of the IBFM wave
functions of the lowest-lying states in $^{147}$Ba is different
from those for $^{143,145}$Ba: they are 
dominated by the $1h_{9/2}$ configuration in $^{143,145}$Ba, whereas in $^{147}$Ba 
low-energy negative parity states are characterized by the mixing of the 
$3p_{1/2,3/2}2f_{5/2,7/2}$ single-particle configurations.

One also notices from Table~\ref{tab:frac-147ba} that states in the lowest
two bands, built on the ${3/2}^-_1$ and ${1/2}^-_1$, do not
contain $f$-boson components in their wave functions. 
The negative-parity bands built on the ${7/2}^-_2$ and ${9/2}^-_1$
states predominantly correspond to the 
$1h_{9/2}$ configuration coupled with the $sd$-boson space, and again
there is no octupole $f$-boson component in these bands. 
The calculation predicts the lowest negative-parity octupole band to be the 
one built on the ${11/2}^-_3$ state, and this band is connected by rather strong E3
transition to the ground state band, e.g., $B(E3; {15/2}^-_3\rightarrow
{9/2}^+_1)=29$ W.u. The experimental band built on the ${11/2}$
state at 573 keV (with tentative assignment of positive parity) has been 
identified as a possible octupole structure \cite{rzacaurban2013}. This band can be compared to the theoretical
sequence with the ${11/2}^+_1$ band-head at excitation energy 855 keV, dominated by the 
$1h_{9/2}$ single-particle orbital coupled to the $sdf$-boson space.

\begin{table}[htb!]
\caption{\label{tab:trans-147ba} Same as in the caption to Table~\ref{tab:trans-143ba},
 but for the $^{147}$Ba nucleus.}
\begin{center}
\begin{tabular}{lc}
\hline\hline
\textrm{$B(E\lambda; J_i^{\pi}\rightarrow J_f^{\pi}$)} &
\textrm{Theory (W.u.)} \\
\hline
$B(E2; {7/2}^-_1 \rightarrow {5/2}^-_1)$ & 13 \\
$B(E2; {9/2}^-_1 \rightarrow {5/2}^-_1)$ & 0.89 \\
$B(E2; {9/2}^-_2 \rightarrow {5/2}^-_1)$ & 104 \\
$B(E2; {11/2}^-_2 \rightarrow {7/2}^-_2)$ & 24 \\
$B(E2; {13/2}^-_1 \rightarrow {9/2}^-_1)$ & 44 \\
$B(E2; {11/2}^+_1 \rightarrow {7/2}^+_1)$ & 0.17 \\
$B(E2; {13/2}^+_1 \rightarrow {9/2}^+_1)$ & 92 \\
$B(E2; {15/2}^+_1 \rightarrow {11/2}^+_1)$ & 48 \\
$B(E2; {17/2}^+_1 \rightarrow {13/2}^+_1)$ & 97 \\
$B(E3; {11/2}^+_1 \rightarrow {5/2}^-_1)$ & 0.037 \\
$B(E3; {11/2}^+_2 \rightarrow {5/2}^-_1)$ & 25 \\
$B(E3; {15/2}^+_1 \rightarrow {9/2}^-_1)$ & 15 \\
%
$B(E3; {15/2}^-_3 \rightarrow {9/2}^+_1)$ & 29 \\
$B(E3; {19/2}^-_1 \rightarrow {13/2}^+_1)$ & 38 \\
\hline\hline
\end{tabular}
\end{center}
\end{table}

\begin{table}[htb!]
\caption{\label{tab:frac-147ba} Same as in the caption to Table~\ref{tab:frac-143ba}, but
 for $^{147}$Ba.}
\begin{center}
\begin{tabular}{cccccccc}
\hline\hline
\textrm{$J^{\pi}$} &
\textrm{$\langle\hat n_f\rangle$} &
\textrm{$3p_{1/2}$} &
\textrm{$3p_{3/2}$} &
\textrm{$2f_{5/2}$} &
\textrm{$2f_{7/2}$} &
\textrm{$1h_{9/2}$} &
\textrm{$1i_{13/2}$} \\
\hline
${3/2}^-_1$ & 0.002 & 14 & 38 & 9 & 39 & 0 & 0 \\
${1/2}^-_1$ & 0.000 & 20 & 35 & 17 & 28 & 0 & 0 \\
${7/2}^-_2$ & 0.000 & 0 & 0 & 0 & 2 & 98 & 0 \\
${9/2}^-_1$ & 0.000 & 0 & 0 & 1 & 2 & 97 & 0 \\
${11/2}^-_3$ & 0.996 & 0 & 0 & 0 & 0 & 0 & 100 \\
${9/2}^+_1$ & 0.002 & 0 & 0 & 0 & 0 & 0 & 100 \\
${5/2}^+_1$ & 0.950 & 10 & 31 & 8 & 45 & 0 & 6 \\
${3/2}^+_1$ & 0.999 & 14 & 33 & 15 & 35 & 3 & 0 \\
${11/2}^+_1$ & 1.000 & 0 & 0 & 1 & 2 & 97 & 0 \\
\hline\hline
\end{tabular}
\end{center}
\end{table}

\section{Conclusions\label{sec:summary}}

The role of octupole correlations and the relevant spectroscopic
properties of neutron-rich odd-mass Ba isotopes have been analyzed in a theoretical framework based on nuclear
density functional theory and the particle-core coupling scheme. 
In the particular method employed in the present study, the interacting-boson Hamiltonian 
that describes the even-even
core nucleus, as well as the single-particle energies and occupation
probabilities of an unpaired nucleon, are completely determined by 
constrained SCMF calculations for a given choice of the energy density functional
and pairing interaction. 
Only the coupling constants for the boson-fermion interaction are 
adjusted to selected spectroscopic data for the low-lying
states in the odd-mass systems.

In this work the $sdf$-IBFM framework has been implemented: the boson-core
Hamiltonian involves both quadrupole and octupole boson degrees of
freedom and is constructed fully microscopically by mapping the axially-symmetric 
($\beta_2, \beta_3$) deformation energy surface obtained by a constrained
relativistic Hartree-Bogoliubov SCMF calculation onto the expectation value of the
Hamiltonian in the $sdf$-boson condensate state. 
In the odd-mass Ba nuclei considered here the role of octupole deformation is not 
very important for the lowest levels near the ground state, and the adjustment of the
boson-fermion strength parameters is relatively straightforward, even though there are
many terms in the corresponding Hamiltonian Eq.~(\ref{eq:bf}).

The SCMF ($\beta_2, \beta_3$) deformation energy surfaces for the even-even
Ba nuclei exhibit a transition from a weakly deformed quadrupole shape of 
$^{142}$Ba to moderately quadrupole and octupole deformed shapes of $^{144,146}$Ba,
characterized by $\beta_3$-soft potentials. The resulting $sdf$ IBM
energy spectra display a signature of octupole collectivity in the pronounced E3 transitions
between the low-lying negative-parity band and the ground-state band, in 
agreement with recent spectroscopic data \cite{bucher2016,bucher2017}. 
The $sdf$-IBFM reproduces the experimental low-energy excitation spectra
in the considered odd-mass Ba isotopes fairly well. 
In particular, the present calculation indicates that octupole correlations  
are not present in the lowest states of $^{143,145,147}$Ba nuclei: most
of their low-lying positive- and negative-parity yrast bands are predominantly 
formed by coupling the odd-neutron orbitals to the $sd$ boson space.  
Octupole states have been identified at somewhat higher excitation energy
-- e.g., in $^{145}$Ba the bands 
built on the ${15/2}^-_4$ and ${7/2}^+_1$ states are characterized by the coupling of 
the odd-neutron to the $sd+f$ boson space, and exhibit pronounced 
E3 transitions to the ground state band. 
These results, especially for $^{145,147}$Ba, are consistent with the 
conclusion of recent experimental studies
\cite{rzacaurban2012,rzacaurban2013}.

A particularly interesting case for a follow-up study are 
spectroscopic properties of actinide
nuclei, e.g., $^{224,225}$Ra, where signatures of stable octupole
shapes have been suggested and identified experimentally, such
as parity doublets, pronounced electric dipole and octupole transitions. 
The low-energy states of these actinide nuclei are much
richer in structure compared to the present case, as octupole correlations are expected 
to be as prominent as the quadrupole ones. 
A quantitative analysis of quadrupole and octupole degrees of freedom in odd-mass nuclei in this 
region certainly presents a challenging 
application of the method introduced in the present work.

\acknowledgments

Part of this work has been completed during the visit of K.N. to the 
Institut f\"ur Kernphysik (IKP), University of Cologne. 
He acknowledges IKP Cologne and Jan Jolie for their kind
hospitality and financial support. 
This work was supported in part by the Croatian Science Foundation --
project "Structure and Dynamics of Exotic Femtosystems" (IP-2014-09-9159)
and the QuantiXLie Centre of Excellence, a project co-financed by the
Croatian Government and European Union through the European Regional
Development Fund - the Competitiveness and Cohesion Operational Programme
(KK.01.1.1.01).

\bibliography{refs}

\begin{thebibliography}{30}%
\makeatletter
\providecommand \@ifxundefined [1]{%
 \@ifx{#1\undefined}
}%
\providecommand \@ifnum [1]{%
 \ifnum #1\expandafter \@firstoftwo
 \else \expandafter \@secondoftwo
 \fi
}%
\providecommand \@ifx [1]{%
 \ifx #1\expandafter \@firstoftwo
 \else \expandafter \@secondoftwo
 \fi
}%
\providecommand \natexlab [1]{#1}%
\providecommand \enquote  [1]{``#1''}%
\providecommand \bibnamefont  [1]{#1}%
\providecommand \bibfnamefont [1]{#1}%
\providecommand \citenamefont [1]{#1}%
\providecommand \href@noop [0]{\@secondoftwo}%
\providecommand \href [0]{\begingroup \@sanitize@url \@href}%
\providecommand \@href[1]{\@@startlink{#1}\@@href}%
\providecommand \@@href[1]{\endgroup#1\@@endlink}%
\providecommand \@sanitize@url [0]{\catcode `\\12\catcode `\$12\catcode
  `\&12\catcode `\#12\catcode `\^12\catcode `\_12\catcode `\%12\relax}%
\providecommand \@@startlink[1]{}%
\providecommand \@@endlink[0]{}%
\providecommand \url  [0]{\begingroup\@sanitize@url \@url }%
\providecommand \@url [1]{\endgroup\@href {#1}{\urlprefix }}%
\providecommand \urlprefix  [0]{URL }%
\providecommand \Eprint [0]{\href }%
\providecommand \doibase [0]{http://dx.doi.org/}%
\providecommand \selectlanguage [0]{\@gobble}%
\providecommand \bibinfo  [0]{\@secondoftwo}%
\providecommand \bibfield  [0]{\@secondoftwo}%
\providecommand \translation [1]{[#1]}%
\providecommand \BibitemOpen [0]{}%
\providecommand \bibitemStop [0]{}%
\providecommand \bibitemNoStop [0]{.\EOS\space}%
\providecommand \EOS [0]{\spacefactor3000\relax}%
\providecommand \BibitemShut  [1]{\csname bibitem#1\endcsname}%
\let\auto@bib@innerbib\@empty
\bibitem [{\citenamefont {Butler}\ and\ \citenamefont
  {Nazarewicz}(1996)}]{butler1996}%
  \BibitemOpen
  \bibfield  {author} {\bibinfo {author} {\bibfnamefont {P.~A.}\ \bibnamefont
  {Butler}}\ and\ \bibinfo {author} {\bibfnamefont {W.}~\bibnamefont
  {Nazarewicz}},\ }\href {\doibase 10.1103/RevModPhys.68.349} {\bibfield
  {journal} {\bibinfo  {journal} {Rev. Mod. Phys.}\ }\textbf {\bibinfo {volume}
  {68}},\ \bibinfo {pages} {349} (\bibinfo {year} {1996})}\BibitemShut
  {NoStop}%
\bibitem [{\citenamefont {Haxton}\ and\ \citenamefont
  {Henley}(1983)}]{haxton1983}%
  \BibitemOpen
  \bibfield  {author} {\bibinfo {author} {\bibfnamefont {W.~C.}\ \bibnamefont
  {Haxton}}\ and\ \bibinfo {author} {\bibfnamefont {E.~M.}\ \bibnamefont
  {Henley}},\ }\href {\doibase 10.1103/PhysRevLett.51.1937} {\bibfield
  {journal} {\bibinfo  {journal} {Phys. Rev. Lett.}\ }\textbf {\bibinfo
  {volume} {51}},\ \bibinfo {pages} {1937} (\bibinfo {year}
  {1983})}\BibitemShut {NoStop}%
\bibitem [{\citenamefont {Dobaczewski}\ and\ \citenamefont
  {Engel}(2005)}]{dobaczewski2005}%
  \BibitemOpen
  \bibfield  {author} {\bibinfo {author} {\bibfnamefont {J.}~\bibnamefont
  {Dobaczewski}}\ and\ \bibinfo {author} {\bibfnamefont {J.}~\bibnamefont
  {Engel}},\ }\href {\doibase 10.1103/PhysRevLett.94.232502} {\bibfield
  {journal} {\bibinfo  {journal} {Phys. Rev. Lett.}\ }\textbf {\bibinfo
  {volume} {94}},\ \bibinfo {pages} {232502} (\bibinfo {year}
  {2005})}\BibitemShut {NoStop}%
\bibitem [{\citenamefont {Gaffney}\ \emph {et~al.}(2013)\citenamefont
  {Gaffney}, \citenamefont {Butler}, \citenamefont {Scheck}, \citenamefont
  {Hayes}, \citenamefont {Wenander}, \citenamefont {Albers}, \citenamefont
  {Bastin}, \citenamefont {Bauer}, \citenamefont {Blazhev}, \citenamefont
  {B\"onig}, \citenamefont {Bree}, \citenamefont {Cederk\"all}, \citenamefont
  {Chupp}, \citenamefont {Cline}, \citenamefont {Cocolios}, \citenamefont
  {Davinson}, \citenamefont {Witte}, \citenamefont {Diriken}, \citenamefont
  {Grahn}, \citenamefont {Herzan}, \citenamefont {Huyse}, \citenamefont
  {Jenkins}, \citenamefont {Joss}, \citenamefont {Kesteloot}, \citenamefont
  {Konki}, \citenamefont {Kowalczyk}, \citenamefont {Kr^^c3^^b6ll},
  \citenamefont {Kwan}, \citenamefont {Lutter}, \citenamefont {Moschner},
  \citenamefont {Napiorkowski}, \citenamefont {Pakarinen}, \citenamefont
  {Pfeiffer}, \citenamefont {Radeck}, \citenamefont {Reiter}, \citenamefont
  {Reynders}, \citenamefont {Rigby}, \citenamefont {Robledo}, \citenamefont
  {Rudigier}, \citenamefont {Sambi}, \citenamefont {Seidlitz}, \citenamefont
  {Siebeck}, \citenamefont {Stora}, \citenamefont {Thoele}, \citenamefont
  {Duppen}, \citenamefont {Vermeulen}, \citenamefont {von Schmid},
  \citenamefont {Voulot}, \citenamefont {Warr}, \citenamefont {Wimmer},
  \citenamefont {Wrzosek-Lipska}, \citenamefont {Wu},\ and\ \citenamefont
  {Zielinska}}]{gaffney2013}%
  \BibitemOpen
  \bibfield  {author} {\bibinfo {author} {\bibfnamefont {L.~P.}\ \bibnamefont
  {Gaffney}}, \bibinfo {author} {\bibfnamefont {P.~A.}\ \bibnamefont {Butler}},
  \bibinfo {author} {\bibfnamefont {M.}~\bibnamefont {Scheck}}, \bibinfo
  {author} {\bibfnamefont {A.~B.}\ \bibnamefont {Hayes}}, \bibinfo {author}
  {\bibfnamefont {F.}~\bibnamefont {Wenander}}, \bibinfo {author}
  {\bibfnamefont {M.}~\bibnamefont {Albers}}, \bibinfo {author} {\bibfnamefont
  {B.}~\bibnamefont {Bastin}}, \bibinfo {author} {\bibfnamefont
  {C.}~\bibnamefont {Bauer}}, \bibinfo {author} {\bibfnamefont
  {A.}~\bibnamefont {Blazhev}}, \bibinfo {author} {\bibfnamefont
  {S.}~\bibnamefont {B\"onig}}, \bibinfo {author} {\bibfnamefont
  {N.}~\bibnamefont {Bree}}, \bibinfo {author} {\bibfnamefont {J.}~\bibnamefont
  {Cederk\"all}}, \bibinfo {author} {\bibfnamefont {T.}~\bibnamefont {Chupp}},
  \bibinfo {author} {\bibfnamefont {D.}~\bibnamefont {Cline}}, \bibinfo
  {author} {\bibfnamefont {T.~E.}\ \bibnamefont {Cocolios}}, \bibinfo {author}
  {\bibfnamefont {T.}~\bibnamefont {Davinson}}, \bibinfo {author}
  {\bibfnamefont {H.~D.}\ \bibnamefont {Witte}}, \bibinfo {author}
  {\bibfnamefont {J.}~\bibnamefont {Diriken}}, \bibinfo {author} {\bibfnamefont
  {T.}~\bibnamefont {Grahn}}, \bibinfo {author} {\bibfnamefont
  {A.}~\bibnamefont {Herzan}}, \bibinfo {author} {\bibfnamefont
  {M.}~\bibnamefont {Huyse}}, \bibinfo {author} {\bibfnamefont {D.~G.}\
  \bibnamefont {Jenkins}}, \bibinfo {author} {\bibfnamefont {D.~T.}\
  \bibnamefont {Joss}}, \bibinfo {author} {\bibfnamefont {N.}~\bibnamefont
  {Kesteloot}}, \bibinfo {author} {\bibfnamefont {J.}~\bibnamefont {Konki}},
  \bibinfo {author} {\bibfnamefont {M.}~\bibnamefont {Kowalczyk}}, \bibinfo
  {author} {\bibfnamefont {T.}~\bibnamefont {Kr^^c3^^b6ll}}, \bibinfo {author}
  {\bibfnamefont {E.}~\bibnamefont {Kwan}}, \bibinfo {author} {\bibfnamefont
  {R.}~\bibnamefont {Lutter}}, \bibinfo {author} {\bibfnamefont
  {K.}~\bibnamefont {Moschner}}, \bibinfo {author} {\bibfnamefont
  {P.}~\bibnamefont {Napiorkowski}}, \bibinfo {author} {\bibfnamefont
  {J.}~\bibnamefont {Pakarinen}}, \bibinfo {author} {\bibfnamefont
  {M.}~\bibnamefont {Pfeiffer}}, \bibinfo {author} {\bibfnamefont
  {D.}~\bibnamefont {Radeck}}, \bibinfo {author} {\bibfnamefont
  {P.}~\bibnamefont {Reiter}}, \bibinfo {author} {\bibfnamefont
  {K.}~\bibnamefont {Reynders}}, \bibinfo {author} {\bibfnamefont {S.~V.}\
  \bibnamefont {Rigby}}, \bibinfo {author} {\bibfnamefont {L.~M.}\ \bibnamefont
  {Robledo}}, \bibinfo {author} {\bibfnamefont {M.}~\bibnamefont {Rudigier}},
  \bibinfo {author} {\bibfnamefont {S.}~\bibnamefont {Sambi}}, \bibinfo
  {author} {\bibfnamefont {M.}~\bibnamefont {Seidlitz}}, \bibinfo {author}
  {\bibfnamefont {B.}~\bibnamefont {Siebeck}}, \bibinfo {author} {\bibfnamefont
  {T.}~\bibnamefont {Stora}}, \bibinfo {author} {\bibfnamefont
  {P.}~\bibnamefont {Thoele}}, \bibinfo {author} {\bibfnamefont {P.~V.}\
  \bibnamefont {Duppen}}, \bibinfo {author} {\bibfnamefont {M.~J.}\
  \bibnamefont {Vermeulen}}, \bibinfo {author} {\bibfnamefont {M.}~\bibnamefont
  {von Schmid}}, \bibinfo {author} {\bibfnamefont {D.}~\bibnamefont {Voulot}},
  \bibinfo {author} {\bibfnamefont {N.}~\bibnamefont {Warr}}, \bibinfo {author}
  {\bibfnamefont {K.}~\bibnamefont {Wimmer}}, \bibinfo {author} {\bibfnamefont
  {K.}~\bibnamefont {Wrzosek-Lipska}}, \bibinfo {author} {\bibfnamefont
  {C.~Y.}\ \bibnamefont {Wu}}, \ and\ \bibinfo {author} {\bibfnamefont
  {M.}~\bibnamefont {Zielinska}},\ }\href {\doibase 10.1038/nature12073}
  {\bibfield  {journal} {\bibinfo  {journal} {Nature (London)}\ }\textbf
  {\bibinfo {volume} {497}},\ \bibinfo {pages} {199} (\bibinfo {year}
  {2013})}\BibitemShut {NoStop}%
\bibitem [{\citenamefont {Bucher}\ \emph {et~al.}(2016)\citenamefont {Bucher},
  \citenamefont {Zhu}, \citenamefont {Wu}, \citenamefont {Janssens},
  \citenamefont {Cline}, \citenamefont {Hayes}, \citenamefont {Albers},
  \citenamefont {Ayangeakaa}, \citenamefont {Butler}, \citenamefont {Campbell},
  \citenamefont {Carpenter}, \citenamefont {Chiara}, \citenamefont {Clark},
  \citenamefont {Crawford}, \citenamefont {Cromaz}, \citenamefont {David},
  \citenamefont {Dickerson}, \citenamefont {Gregor}, \citenamefont {Harker},
  \citenamefont {Hoffman}, \citenamefont {Kay}, \citenamefont {Kondev},
  \citenamefont {Korichi}, \citenamefont {Lauritsen}, \citenamefont
  {Macchiavelli}, \citenamefont {Pardo}, \citenamefont {Richard}, \citenamefont
  {Riley}, \citenamefont {Savard}, \citenamefont {Scheck}, \citenamefont
  {Seweryniak}, \citenamefont {Smith}, \citenamefont {Vondrasek},\ and\
  \citenamefont {Wiens}}]{bucher2016}%
  \BibitemOpen
  \bibfield  {author} {\bibinfo {author} {\bibfnamefont {B.}~\bibnamefont
  {Bucher}}, \bibinfo {author} {\bibfnamefont {S.}~\bibnamefont {Zhu}},
  \bibinfo {author} {\bibfnamefont {C.~Y.}\ \bibnamefont {Wu}}, \bibinfo
  {author} {\bibfnamefont {R.~V.~F.}\ \bibnamefont {Janssens}}, \bibinfo
  {author} {\bibfnamefont {D.}~\bibnamefont {Cline}}, \bibinfo {author}
  {\bibfnamefont {A.~B.}\ \bibnamefont {Hayes}}, \bibinfo {author}
  {\bibfnamefont {M.}~\bibnamefont {Albers}}, \bibinfo {author} {\bibfnamefont
  {A.~D.}\ \bibnamefont {Ayangeakaa}}, \bibinfo {author} {\bibfnamefont
  {P.~A.}\ \bibnamefont {Butler}}, \bibinfo {author} {\bibfnamefont {C.~M.}\
  \bibnamefont {Campbell}}, \bibinfo {author} {\bibfnamefont {M.~P.}\
  \bibnamefont {Carpenter}}, \bibinfo {author} {\bibfnamefont {C.~J.}\
  \bibnamefont {Chiara}}, \bibinfo {author} {\bibfnamefont {J.~A.}\
  \bibnamefont {Clark}}, \bibinfo {author} {\bibfnamefont {H.~L.}\ \bibnamefont
  {Crawford}}, \bibinfo {author} {\bibfnamefont {M.}~\bibnamefont {Cromaz}},
  \bibinfo {author} {\bibfnamefont {H.~M.}\ \bibnamefont {David}}, \bibinfo
  {author} {\bibfnamefont {C.}~\bibnamefont {Dickerson}}, \bibinfo {author}
  {\bibfnamefont {E.~T.}\ \bibnamefont {Gregor}}, \bibinfo {author}
  {\bibfnamefont {J.}~\bibnamefont {Harker}}, \bibinfo {author} {\bibfnamefont
  {C.~R.}\ \bibnamefont {Hoffman}}, \bibinfo {author} {\bibfnamefont {B.~P.}\
  \bibnamefont {Kay}}, \bibinfo {author} {\bibfnamefont {F.~G.}\ \bibnamefont
  {Kondev}}, \bibinfo {author} {\bibfnamefont {A.}~\bibnamefont {Korichi}},
  \bibinfo {author} {\bibfnamefont {T.}~\bibnamefont {Lauritsen}}, \bibinfo
  {author} {\bibfnamefont {A.~O.}\ \bibnamefont {Macchiavelli}}, \bibinfo
  {author} {\bibfnamefont {R.~C.}\ \bibnamefont {Pardo}}, \bibinfo {author}
  {\bibfnamefont {A.}~\bibnamefont {Richard}}, \bibinfo {author} {\bibfnamefont
  {M.~A.}\ \bibnamefont {Riley}}, \bibinfo {author} {\bibfnamefont
  {G.}~\bibnamefont {Savard}}, \bibinfo {author} {\bibfnamefont
  {M.}~\bibnamefont {Scheck}}, \bibinfo {author} {\bibfnamefont
  {D.}~\bibnamefont {Seweryniak}}, \bibinfo {author} {\bibfnamefont {M.~K.}\
  \bibnamefont {Smith}}, \bibinfo {author} {\bibfnamefont {R.}~\bibnamefont
  {Vondrasek}}, \ and\ \bibinfo {author} {\bibfnamefont {A.}~\bibnamefont
  {Wiens}},\ }\href {\doibase 10.1103/PhysRevLett.116.112503} {\bibfield
  {journal} {\bibinfo  {journal} {Phys. Rev. Lett.}\ }\textbf {\bibinfo
  {volume} {116}},\ \bibinfo {pages} {112503} (\bibinfo {year}
  {2016})}\BibitemShut {NoStop}%
\bibitem [{\citenamefont {Bucher}\ \emph {et~al.}(2017)\citenamefont {Bucher},
  \citenamefont {Zhu}, \citenamefont {Wu}, \citenamefont {Janssens},
  \citenamefont {Bernard}, \citenamefont {Robledo}, \citenamefont
  {Rodr\'{\i}guez}, \citenamefont {Cline}, \citenamefont {Hayes}, \citenamefont
  {Ayangeakaa}, \citenamefont {Buckner}, \citenamefont {Campbell},
  \citenamefont {Carpenter}, \citenamefont {Clark}, \citenamefont {Crawford},
  \citenamefont {David}, \citenamefont {Dickerson}, \citenamefont {Harker},
  \citenamefont {Hoffman}, \citenamefont {Kay}, \citenamefont {Kondev},
  \citenamefont {Lauritsen}, \citenamefont {Macchiavelli}, \citenamefont
  {Pardo}, \citenamefont {Savard}, \citenamefont {Seweryniak},\ and\
  \citenamefont {Vondrasek}}]{bucher2017}%
  \BibitemOpen
  \bibfield  {author} {\bibinfo {author} {\bibfnamefont {B.}~\bibnamefont
  {Bucher}}, \bibinfo {author} {\bibfnamefont {S.}~\bibnamefont {Zhu}},
  \bibinfo {author} {\bibfnamefont {C.~Y.}\ \bibnamefont {Wu}}, \bibinfo
  {author} {\bibfnamefont {R.~V.~F.}\ \bibnamefont {Janssens}}, \bibinfo
  {author} {\bibfnamefont {R.~N.}\ \bibnamefont {Bernard}}, \bibinfo {author}
  {\bibfnamefont {L.~M.}\ \bibnamefont {Robledo}}, \bibinfo {author}
  {\bibfnamefont {T.~R.}\ \bibnamefont {Rodr\'{\i}guez}}, \bibinfo {author}
  {\bibfnamefont {D.}~\bibnamefont {Cline}}, \bibinfo {author} {\bibfnamefont
  {A.~B.}\ \bibnamefont {Hayes}}, \bibinfo {author} {\bibfnamefont {A.~D.}\
  \bibnamefont {Ayangeakaa}}, \bibinfo {author} {\bibfnamefont {M.~Q.}\
  \bibnamefont {Buckner}}, \bibinfo {author} {\bibfnamefont {C.~M.}\
  \bibnamefont {Campbell}}, \bibinfo {author} {\bibfnamefont {M.~P.}\
  \bibnamefont {Carpenter}}, \bibinfo {author} {\bibfnamefont {J.~A.}\
  \bibnamefont {Clark}}, \bibinfo {author} {\bibfnamefont {H.~L.}\ \bibnamefont
  {Crawford}}, \bibinfo {author} {\bibfnamefont {H.~M.}\ \bibnamefont {David}},
  \bibinfo {author} {\bibfnamefont {C.}~\bibnamefont {Dickerson}}, \bibinfo
  {author} {\bibfnamefont {J.}~\bibnamefont {Harker}}, \bibinfo {author}
  {\bibfnamefont {C.~R.}\ \bibnamefont {Hoffman}}, \bibinfo {author}
  {\bibfnamefont {B.~P.}\ \bibnamefont {Kay}}, \bibinfo {author} {\bibfnamefont
  {F.~G.}\ \bibnamefont {Kondev}}, \bibinfo {author} {\bibfnamefont
  {T.}~\bibnamefont {Lauritsen}}, \bibinfo {author} {\bibfnamefont {A.~O.}\
  \bibnamefont {Macchiavelli}}, \bibinfo {author} {\bibfnamefont {R.~C.}\
  \bibnamefont {Pardo}}, \bibinfo {author} {\bibfnamefont {G.}~\bibnamefont
  {Savard}}, \bibinfo {author} {\bibfnamefont {D.}~\bibnamefont {Seweryniak}},
  \ and\ \bibinfo {author} {\bibfnamefont {R.}~\bibnamefont {Vondrasek}},\
  }\href {\doibase 10.1103/PhysRevLett.118.152504} {\bibfield  {journal}
  {\bibinfo  {journal} {Phys. Rev. Lett.}\ }\textbf {\bibinfo {volume} {118}},\
  \bibinfo {pages} {152504} (\bibinfo {year} {2017})}\BibitemShut {NoStop}%
\bibitem [{\citenamefont {Parker}\ \emph {et~al.}(2015)\citenamefont {Parker},
  \citenamefont {Dietrich}, \citenamefont {Kalita}, \citenamefont {Lemke},
  \citenamefont {Bailey}, \citenamefont {Bishof}, \citenamefont {Greene},
  \citenamefont {Holt}, \citenamefont {Korsch}, \citenamefont {Lu},
  \citenamefont {Mueller}, \citenamefont {O'Connor},\ and\ \citenamefont
  {Singh}}]{parker2015}%
  \BibitemOpen
  \bibfield  {author} {\bibinfo {author} {\bibfnamefont {R.~H.}\ \bibnamefont
  {Parker}}, \bibinfo {author} {\bibfnamefont {M.~R.}\ \bibnamefont
  {Dietrich}}, \bibinfo {author} {\bibfnamefont {M.~R.}\ \bibnamefont
  {Kalita}}, \bibinfo {author} {\bibfnamefont {N.~D.}\ \bibnamefont {Lemke}},
  \bibinfo {author} {\bibfnamefont {K.~G.}\ \bibnamefont {Bailey}}, \bibinfo
  {author} {\bibfnamefont {M.}~\bibnamefont {Bishof}}, \bibinfo {author}
  {\bibfnamefont {J.~P.}\ \bibnamefont {Greene}}, \bibinfo {author}
  {\bibfnamefont {R.~J.}\ \bibnamefont {Holt}}, \bibinfo {author}
  {\bibfnamefont {W.}~\bibnamefont {Korsch}}, \bibinfo {author} {\bibfnamefont
  {Z.-T.}\ \bibnamefont {Lu}}, \bibinfo {author} {\bibfnamefont
  {P.}~\bibnamefont {Mueller}}, \bibinfo {author} {\bibfnamefont {T.~P.}\
  \bibnamefont {O'Connor}}, \ and\ \bibinfo {author} {\bibfnamefont {J.~T.}\
  \bibnamefont {Singh}},\ }\href {\doibase 10.1103/PhysRevLett.114.233002}
  {\bibfield  {journal} {\bibinfo  {journal} {Phys. Rev. Lett.}\ }\textbf
  {\bibinfo {volume} {114}},\ \bibinfo {pages} {233002} (\bibinfo {year}
  {2015})}\BibitemShut {NoStop}%
\bibitem [{\citenamefont {Griffith}\ \emph {et~al.}(2009)\citenamefont
  {Griffith}, \citenamefont {Swallows}, \citenamefont {Loftus}, \citenamefont
  {Romalis}, \citenamefont {Heckel},\ and\ \citenamefont
  {Fortson}}]{griffis1983}%
  \BibitemOpen
  \bibfield  {author} {\bibinfo {author} {\bibfnamefont {W.~C.}\ \bibnamefont
  {Griffith}}, \bibinfo {author} {\bibfnamefont {M.~D.}\ \bibnamefont
  {Swallows}}, \bibinfo {author} {\bibfnamefont {T.~H.}\ \bibnamefont
  {Loftus}}, \bibinfo {author} {\bibfnamefont {M.~V.}\ \bibnamefont {Romalis}},
  \bibinfo {author} {\bibfnamefont {B.~R.}\ \bibnamefont {Heckel}}, \ and\
  \bibinfo {author} {\bibfnamefont {E.~N.}\ \bibnamefont {Fortson}},\ }\href
  {\doibase 10.1103/PhysRevLett.102.101601} {\bibfield  {journal} {\bibinfo
  {journal} {Phys. Rev. Lett.}\ }\textbf {\bibinfo {volume} {102}},\ \bibinfo
  {pages} {101601} (\bibinfo {year} {2009})}\BibitemShut {NoStop}%
\bibitem [{\citenamefont {Nomura}\ \emph {et~al.}(2016)\citenamefont {Nomura},
  \citenamefont {Nik\ifmmode \check{s}\else \v{s}\fi{}i\ifmmode~\acute{c}\else
  \'{c}\fi{}},\ and\ \citenamefont {Vretenar}}]{nomura2016odd}%
  \BibitemOpen
  \bibfield  {author} {\bibinfo {author} {\bibfnamefont {K.}~\bibnamefont
  {Nomura}}, \bibinfo {author} {\bibfnamefont {T.}~\bibnamefont {Nik\ifmmode
  \check{s}\else \v{s}\fi{}i\ifmmode~\acute{c}\else \'{c}\fi{}}}, \ and\
  \bibinfo {author} {\bibfnamefont {D.}~\bibnamefont {Vretenar}},\ }\href@noop
  {} {\bibfield  {journal} {\bibinfo  {journal} {Phys. Rev. C}\ }\textbf
  {\bibinfo {volume} {93}},\ \bibinfo {pages} {054305} (\bibinfo {year}
  {2016})}\BibitemShut {NoStop}%
\bibitem [{\citenamefont {Iachello}\ and\ \citenamefont {Arima}(1987)}]{IBM}%
  \BibitemOpen
  \bibfield  {author} {\bibinfo {author} {\bibfnamefont {F.}~\bibnamefont
  {Iachello}}\ and\ \bibinfo {author} {\bibfnamefont {A.}~\bibnamefont
  {Arima}},\ }\href@noop {} {\emph {\bibinfo {title} {The interacting boson
  model}}}\ (\bibinfo  {publisher} {Cambridge University Press, Cambridge},\
  \bibinfo {year} {1987})\BibitemShut {NoStop}%
\bibitem [{\citenamefont {Iachello}\ and\ \citenamefont {{Van
  Isacker}}(1991)}]{IBFM}%
  \BibitemOpen
  \bibfield  {author} {\bibinfo {author} {\bibfnamefont {F.}~\bibnamefont
  {Iachello}}\ and\ \bibinfo {author} {\bibfnamefont {P.}~\bibnamefont {{Van
  Isacker}}},\ }\href@noop {} {\emph {\bibinfo {title} {The interacting
  boson-fermion model}}}\ (\bibinfo  {publisher} {Cambridge University Press,
  Cambridge},\ \bibinfo {year} {1991})\BibitemShut {NoStop}%
\bibitem [{\citenamefont {Zhu}\ \emph {et~al.}(1999)\citenamefont {Zhu},
  \citenamefont {Hamilton}, \citenamefont {Ramayya}, \citenamefont {Jones},
  \citenamefont {Hwang}, \citenamefont {Wang}, \citenamefont {Zhang},
  \citenamefont {Gore}, \citenamefont {Peker}, \citenamefont {Drafta},
  \citenamefont {Babu}, \citenamefont {Ma}, \citenamefont {Long}, \citenamefont
  {Zhu}, \citenamefont {Gan}, \citenamefont {Yang}, \citenamefont {Sakhaee},
  \citenamefont {Li}, \citenamefont {Deng}, \citenamefont {Ginter},
  \citenamefont {Beyer}, \citenamefont {Kormicki}, \citenamefont {Cole},
  \citenamefont {Aryaeinejad}, \citenamefont {Drigert}, \citenamefont
  {Rasmussen}, \citenamefont {Asztalos}, \citenamefont {Lee}, \citenamefont
  {Macchiavelli}, \citenamefont {Chu}, \citenamefont {Gregorich}, \citenamefont
  {Mohar}, \citenamefont {Ter-Akopian}, \citenamefont {Daniel}, \citenamefont
  {Oganessian}, \citenamefont {Donangelo}, \citenamefont {Stoyer},
  \citenamefont {Lougheed}, \citenamefont {Moody}, \citenamefont {Wild},
  \citenamefont {Prussin}, \citenamefont {Kliman},\ and\ \citenamefont
  {Griffin}}]{zhu1999}%
  \BibitemOpen
  \bibfield  {author} {\bibinfo {author} {\bibfnamefont {S.~J.}\ \bibnamefont
  {Zhu}}, \bibinfo {author} {\bibfnamefont {J.~H.}\ \bibnamefont {Hamilton}},
  \bibinfo {author} {\bibfnamefont {A.~V.}\ \bibnamefont {Ramayya}}, \bibinfo
  {author} {\bibfnamefont {E.~F.}\ \bibnamefont {Jones}}, \bibinfo {author}
  {\bibfnamefont {J.~K.}\ \bibnamefont {Hwang}}, \bibinfo {author}
  {\bibfnamefont {M.~G.}\ \bibnamefont {Wang}}, \bibinfo {author}
  {\bibfnamefont {X.~Q.}\ \bibnamefont {Zhang}}, \bibinfo {author}
  {\bibfnamefont {P.~M.}\ \bibnamefont {Gore}}, \bibinfo {author}
  {\bibfnamefont {L.~K.}\ \bibnamefont {Peker}}, \bibinfo {author}
  {\bibfnamefont {G.}~\bibnamefont {Drafta}}, \bibinfo {author} {\bibfnamefont
  {B.~R.~S.}\ \bibnamefont {Babu}}, \bibinfo {author} {\bibfnamefont {W.~C.}\
  \bibnamefont {Ma}}, \bibinfo {author} {\bibfnamefont {G.~L.}\ \bibnamefont
  {Long}}, \bibinfo {author} {\bibfnamefont {L.~Y.}\ \bibnamefont {Zhu}},
  \bibinfo {author} {\bibfnamefont {C.~Y.}\ \bibnamefont {Gan}}, \bibinfo
  {author} {\bibfnamefont {L.~M.}\ \bibnamefont {Yang}}, \bibinfo {author}
  {\bibfnamefont {M.}~\bibnamefont {Sakhaee}}, \bibinfo {author} {\bibfnamefont
  {M.}~\bibnamefont {Li}}, \bibinfo {author} {\bibfnamefont {J.~K.}\
  \bibnamefont {Deng}}, \bibinfo {author} {\bibfnamefont {T.~N.}\ \bibnamefont
  {Ginter}}, \bibinfo {author} {\bibfnamefont {C.~J.}\ \bibnamefont {Beyer}},
  \bibinfo {author} {\bibfnamefont {J.}~\bibnamefont {Kormicki}}, \bibinfo
  {author} {\bibfnamefont {J.~D.}\ \bibnamefont {Cole}}, \bibinfo {author}
  {\bibfnamefont {R.}~\bibnamefont {Aryaeinejad}}, \bibinfo {author}
  {\bibfnamefont {M.~W.}\ \bibnamefont {Drigert}}, \bibinfo {author}
  {\bibfnamefont {J.~O.}\ \bibnamefont {Rasmussen}}, \bibinfo {author}
  {\bibfnamefont {S.}~\bibnamefont {Asztalos}}, \bibinfo {author}
  {\bibfnamefont {I.~Y.}\ \bibnamefont {Lee}}, \bibinfo {author} {\bibfnamefont
  {A.~O.}\ \bibnamefont {Macchiavelli}}, \bibinfo {author} {\bibfnamefont
  {S.~Y.}\ \bibnamefont {Chu}}, \bibinfo {author} {\bibfnamefont {K.~E.}\
  \bibnamefont {Gregorich}}, \bibinfo {author} {\bibfnamefont {M.~F.}\
  \bibnamefont {Mohar}}, \bibinfo {author} {\bibfnamefont {G.~M.}\ \bibnamefont
  {Ter-Akopian}}, \bibinfo {author} {\bibfnamefont {A.~V.}\ \bibnamefont
  {Daniel}}, \bibinfo {author} {\bibfnamefont {Y.~T.}\ \bibnamefont
  {Oganessian}}, \bibinfo {author} {\bibfnamefont {R.}~\bibnamefont
  {Donangelo}}, \bibinfo {author} {\bibfnamefont {M.~A.}\ \bibnamefont
  {Stoyer}}, \bibinfo {author} {\bibfnamefont {R.~W.}\ \bibnamefont
  {Lougheed}}, \bibinfo {author} {\bibfnamefont {K.~J.}\ \bibnamefont {Moody}},
  \bibinfo {author} {\bibfnamefont {J.~F.}\ \bibnamefont {Wild}}, \bibinfo
  {author} {\bibfnamefont {S.~G.}\ \bibnamefont {Prussin}}, \bibinfo {author}
  {\bibfnamefont {J.}~\bibnamefont {Kliman}}, \ and\ \bibinfo {author}
  {\bibfnamefont {H.~C.}\ \bibnamefont {Griffin}},\ }\href {\doibase
  10.1103/PhysRevC.60.051304} {\bibfield  {journal} {\bibinfo  {journal} {Phys.
  Rev. C}\ }\textbf {\bibinfo {volume} {60}},\ \bibinfo {pages} {051304}
  (\bibinfo {year} {1999})}\BibitemShut {NoStop}%
\bibitem [{\citenamefont {Rzaca-Urban}\ \emph {et~al.}(2012)\citenamefont
  {Rzaca-Urban}, \citenamefont {Urban}, \citenamefont {Pinston}, \citenamefont
  {Simpson}, \citenamefont {Smith},\ and\ \citenamefont
  {Ahmad}}]{rzacaurban2012}%
  \BibitemOpen
  \bibfield  {author} {\bibinfo {author} {\bibfnamefont {T.}~\bibnamefont
  {Rzaca-Urban}}, \bibinfo {author} {\bibfnamefont {W.}~\bibnamefont {Urban}},
  \bibinfo {author} {\bibfnamefont {J.~A.}\ \bibnamefont {Pinston}}, \bibinfo
  {author} {\bibfnamefont {G.~S.}\ \bibnamefont {Simpson}}, \bibinfo {author}
  {\bibfnamefont {A.~G.}\ \bibnamefont {Smith}}, \ and\ \bibinfo {author}
  {\bibfnamefont {I.}~\bibnamefont {Ahmad}},\ }\href {\doibase
  10.1103/PhysRevC.86.044324} {\bibfield  {journal} {\bibinfo  {journal} {Phys.
  Rev. C}\ }\textbf {\bibinfo {volume} {86}},\ \bibinfo {pages} {044324}
  (\bibinfo {year} {2012})}\BibitemShut {NoStop}%
\bibitem [{\citenamefont {Leander}\ \emph {et~al.}(1985)\citenamefont
  {Leander}, \citenamefont {Nazarewicz}, \citenamefont {Olanders},
  \citenamefont {Ragnarsson},\ and\ \citenamefont {Dudek}}]{leander1985}%
  \BibitemOpen
  \bibfield  {author} {\bibinfo {author} {\bibfnamefont {G.}~\bibnamefont
  {Leander}}, \bibinfo {author} {\bibfnamefont {W.}~\bibnamefont {Nazarewicz}},
  \bibinfo {author} {\bibfnamefont {P.}~\bibnamefont {Olanders}}, \bibinfo
  {author} {\bibfnamefont {I.}~\bibnamefont {Ragnarsson}}, \ and\ \bibinfo
  {author} {\bibfnamefont {J.}~\bibnamefont {Dudek}},\ }\href {\doibase
  https://doi.org/10.1016/0370-2693(85)90496-4} {\bibfield  {journal} {\bibinfo
   {journal} {Physics Letters B}\ }\textbf {\bibinfo {volume} {152}},\ \bibinfo
  {pages} {284 } (\bibinfo {year} {1985})}\BibitemShut {NoStop}%
\bibitem [{\citenamefont {Nomura}\ \emph {et~al.}(2013)\citenamefont {Nomura},
  \citenamefont {Vretenar},\ and\ \citenamefont {Lu}}]{nomura2013oct}%
  \BibitemOpen
  \bibfield  {author} {\bibinfo {author} {\bibfnamefont {K.}~\bibnamefont
  {Nomura}}, \bibinfo {author} {\bibfnamefont {D.}~\bibnamefont {Vretenar}}, \
  and\ \bibinfo {author} {\bibfnamefont {B.-N.}\ \bibnamefont {Lu}},\ }\href
  {\doibase 10.1103/PhysRevC.88.021303} {\bibfield  {journal} {\bibinfo
  {journal} {Phys. Rev. C}\ }\textbf {\bibinfo {volume} {88}},\ \bibinfo
  {pages} {021303} (\bibinfo {year} {2013})}\BibitemShut {NoStop}%
\bibitem [{\citenamefont {Nomura}\ \emph {et~al.}(2014)\citenamefont {Nomura},
  \citenamefont {Vretenar}, \citenamefont {Nik\ifmmode \check{s}\else
  \v{s}\fi{}i\ifmmode~\acute{c}\else \'{c}\fi{}},\ and\ \citenamefont
  {Lu}}]{nomura2014}%
  \BibitemOpen
  \bibfield  {author} {\bibinfo {author} {\bibfnamefont {K.}~\bibnamefont
  {Nomura}}, \bibinfo {author} {\bibfnamefont {D.}~\bibnamefont {Vretenar}},
  \bibinfo {author} {\bibfnamefont {T.}~\bibnamefont {Nik\ifmmode
  \check{s}\else \v{s}\fi{}i\ifmmode~\acute{c}\else \'{c}\fi{}}}, \ and\
  \bibinfo {author} {\bibfnamefont {B.-N.}\ \bibnamefont {Lu}},\ }\href
  {\doibase 10.1103/PhysRevC.89.024312} {\bibfield  {journal} {\bibinfo
  {journal} {Phys. Rev. C}\ }\textbf {\bibinfo {volume} {89}},\ \bibinfo
  {pages} {024312} (\bibinfo {year} {2014})}\BibitemShut {NoStop}%
\bibitem [{\citenamefont {Chuu}\ \emph {et~al.}(1993)\citenamefont {Chuu},
  \citenamefont {Hsieh},\ and\ \citenamefont {Chiang}}]{chuu1993}%
  \BibitemOpen
  \bibfield  {author} {\bibinfo {author} {\bibfnamefont {D.-S.}\ \bibnamefont
  {Chuu}}, \bibinfo {author} {\bibfnamefont {S.~T.}\ \bibnamefont {Hsieh}}, \
  and\ \bibinfo {author} {\bibfnamefont {H.~C.}\ \bibnamefont {Chiang}},\
  }\href {\doibase 10.1103/PhysRevC.47.183} {\bibfield  {journal} {\bibinfo
  {journal} {Phys. Rev. C}\ }\textbf {\bibinfo {volume} {47}},\ \bibinfo
  {pages} {183} (\bibinfo {year} {1993})}\BibitemShut {NoStop}%
\bibitem [{\citenamefont {Alonso}\ \emph {et~al.}(1995)\citenamefont {Alonso},
  \citenamefont {Arias}, \citenamefont {Frank}, \citenamefont {Sofia},
  \citenamefont {Lenzi},\ and\ \citenamefont {Vitturi}}]{alonso1995}%
  \BibitemOpen
  \bibfield  {author} {\bibinfo {author} {\bibfnamefont {C.}~\bibnamefont
  {Alonso}}, \bibinfo {author} {\bibfnamefont {J.}~\bibnamefont {Arias}},
  \bibinfo {author} {\bibfnamefont {A.}~\bibnamefont {Frank}}, \bibinfo
  {author} {\bibfnamefont {H.}~\bibnamefont {Sofia}}, \bibinfo {author}
  {\bibfnamefont {S.}~\bibnamefont {Lenzi}}, \ and\ \bibinfo {author}
  {\bibfnamefont {A.}~\bibnamefont {Vitturi}},\ }\href {\doibase
  https://doi.org/10.1016/0375-9474(94)00794-N} {\bibfield  {journal} {\bibinfo
   {journal} {Nuclear Physics A}\ }\textbf {\bibinfo {volume} {586}},\ \bibinfo
  {pages} {100 } (\bibinfo {year} {1995})}\BibitemShut {NoStop}%
\bibitem [{\citenamefont {Singh}\ \emph {et~al.}(1998)\citenamefont {Singh},
  \citenamefont {Gangopadhyay}, \citenamefont {Banerjee}, \citenamefont
  {Bhattacharya}, \citenamefont {Bhowmik}, \citenamefont {Muralithar},
  \citenamefont {Singh}, \citenamefont {Mukherjee}, \citenamefont
  {Datta~Pramanik}, \citenamefont {Goswami}, \citenamefont {Chattopadhyay},
  \citenamefont {Bhattacharya}, \citenamefont {Dasmahapatra},\ and\
  \citenamefont {Sen}}]{singh1998}%
  \BibitemOpen
  \bibfield  {author} {\bibinfo {author} {\bibfnamefont {A.~K.}\ \bibnamefont
  {Singh}}, \bibinfo {author} {\bibfnamefont {G.}~\bibnamefont {Gangopadhyay}},
  \bibinfo {author} {\bibfnamefont {D.}~\bibnamefont {Banerjee}}, \bibinfo
  {author} {\bibfnamefont {R.}~\bibnamefont {Bhattacharya}}, \bibinfo {author}
  {\bibfnamefont {R.~K.}\ \bibnamefont {Bhowmik}}, \bibinfo {author}
  {\bibfnamefont {S.}~\bibnamefont {Muralithar}}, \bibinfo {author}
  {\bibfnamefont {R.~P.}\ \bibnamefont {Singh}}, \bibinfo {author}
  {\bibfnamefont {A.}~\bibnamefont {Mukherjee}}, \bibinfo {author}
  {\bibfnamefont {U.}~\bibnamefont {Datta~Pramanik}}, \bibinfo {author}
  {\bibfnamefont {A.}~\bibnamefont {Goswami}}, \bibinfo {author} {\bibfnamefont
  {S.}~\bibnamefont {Chattopadhyay}}, \bibinfo {author} {\bibfnamefont
  {S.}~\bibnamefont {Bhattacharya}}, \bibinfo {author} {\bibfnamefont
  {B.}~\bibnamefont {Dasmahapatra}}, \ and\ \bibinfo {author} {\bibfnamefont
  {S.}~\bibnamefont {Sen}},\ }\href {\doibase 10.1103/PhysRevC.57.1617}
  {\bibfield  {journal} {\bibinfo  {journal} {Phys. Rev. C}\ }\textbf {\bibinfo
  {volume} {57}},\ \bibinfo {pages} {1617} (\bibinfo {year}
  {1998})}\BibitemShut {NoStop}%
\bibitem [{\citenamefont {Vretenar}\ \emph {et~al.}(2005)\citenamefont
  {Vretenar}, \citenamefont {Afanasjev}, \citenamefont {Lalazissis},\ and\
  \citenamefont {Ring}}]{vretenar2005}%
  \BibitemOpen
  \bibfield  {author} {\bibinfo {author} {\bibfnamefont {D.}~\bibnamefont
  {Vretenar}}, \bibinfo {author} {\bibfnamefont {A.}~\bibnamefont {Afanasjev}},
  \bibinfo {author} {\bibfnamefont {G.}~\bibnamefont {Lalazissis}}, \ and\
  \bibinfo {author} {\bibfnamefont {P.}~\bibnamefont {Ring}},\ }\href {\doibase
  10.1016/j.physrep.2004.10.001} {\bibfield  {journal} {\bibinfo  {journal}
  {Phys. Rep.}\ }\textbf {\bibinfo {volume} {409}},\ \bibinfo {pages} {101 }
  (\bibinfo {year} {2005})}\BibitemShut {NoStop}%
\bibitem [{\citenamefont {Nik\ifmmode \check{s}\else
  \v{s}\fi{}i\ifmmode~\acute{c}\else \'{c}\fi{}}\ \emph
  {et~al.}(2008)\citenamefont {Nik\ifmmode \check{s}\else
  \v{s}\fi{}i\ifmmode~\acute{c}\else \'{c}\fi{}}, \citenamefont {Vretenar},\
  and\ \citenamefont {Ring}}]{DDPC1}%
  \BibitemOpen
  \bibfield  {author} {\bibinfo {author} {\bibfnamefont {T.}~\bibnamefont
  {Nik\ifmmode \check{s}\else \v{s}\fi{}i\ifmmode~\acute{c}\else \'{c}\fi{}}},
  \bibinfo {author} {\bibfnamefont {D.}~\bibnamefont {Vretenar}}, \ and\
  \bibinfo {author} {\bibfnamefont {P.}~\bibnamefont {Ring}},\ }\href {\doibase
  10.1103/PhysRevC.78.034318} {\bibfield  {journal} {\bibinfo  {journal} {Phys.
  Rev. C}\ }\textbf {\bibinfo {volume} {78}},\ \bibinfo {pages} {034318}
  (\bibinfo {year} {2008})}\BibitemShut {NoStop}%
\bibitem [{\citenamefont {Tian}\ \emph {et~al.}(2009)\citenamefont {Tian},
  \citenamefont {Ma},\ and\ \citenamefont {Ring}}]{tian2009}%
  \BibitemOpen
  \bibfield  {author} {\bibinfo {author} {\bibfnamefont {Y.}~\bibnamefont
  {Tian}}, \bibinfo {author} {\bibfnamefont {Z.~Y.}\ \bibnamefont {Ma}}, \ and\
  \bibinfo {author} {\bibfnamefont {P.}~\bibnamefont {Ring}},\ }\href {\doibase
  10.1016/j.physletb.2009.04.067} {\bibfield  {journal} {\bibinfo  {journal}
  {Phys. Lett. B}\ }\textbf {\bibinfo {volume} {676}},\ \bibinfo {pages} {44 }
  (\bibinfo {year} {2009})}\BibitemShut {NoStop}%
\bibitem [{\citenamefont {Barfield}\ \emph {et~al.}(1988)\citenamefont
  {Barfield}, \citenamefont {Barrett}, \citenamefont {Wood},\ and\
  \citenamefont {Scholten}}]{barfield1988}%
  \BibitemOpen
  \bibfield  {author} {\bibinfo {author} {\bibfnamefont {A.~F.}\ \bibnamefont
  {Barfield}}, \bibinfo {author} {\bibfnamefont {B.~R.}\ \bibnamefont
  {Barrett}}, \bibinfo {author} {\bibfnamefont {J.~L.}\ \bibnamefont {Wood}}, \
  and\ \bibinfo {author} {\bibfnamefont {O.}~\bibnamefont {Scholten}},\ }\href
  {\doibase 10.1016/0003-4916(88)90016-4} {\bibfield  {journal} {\bibinfo
  {journal} {Ann. Phys.}\ }\textbf {\bibinfo {volume} {182}},\ \bibinfo {pages}
  {344 } (\bibinfo {year} {1988})}\BibitemShut {NoStop}%
\bibitem [{\citenamefont {Ginocchio}\ and\ \citenamefont
  {Kirson}(1980)}]{ginocchio1980}%
  \BibitemOpen
  \bibfield  {author} {\bibinfo {author} {\bibfnamefont {J.~N.}\ \bibnamefont
  {Ginocchio}}\ and\ \bibinfo {author} {\bibfnamefont {M.~W.}\ \bibnamefont
  {Kirson}},\ }\href {\doibase 10.1016/0375-9474(80)90387-5} {\bibfield
  {journal} {\bibinfo  {journal} {Nucl. Phys. A}\ }\textbf {\bibinfo {volume}
  {350}},\ \bibinfo {pages} {31} (\bibinfo {year} {1980})}\BibitemShut
  {NoStop}%
\bibitem [{\citenamefont {Schaaser}\ and\ \citenamefont
  {Brink}(1986)}]{schaaser1986}%
  \BibitemOpen
  \bibfield  {author} {\bibinfo {author} {\bibfnamefont {H.}~\bibnamefont
  {Schaaser}}\ and\ \bibinfo {author} {\bibfnamefont {D.~M.}\ \bibnamefont
  {Brink}},\ }\href@noop {} {\bibfield  {journal} {\bibinfo  {journal} {Nucl.
  Phys. A}\ }\textbf {\bibinfo {volume} {452}},\ \bibinfo {pages} {1 }
  (\bibinfo {year} {1986})}\BibitemShut {NoStop}%
\bibitem [{\citenamefont {Nomura}\ \emph {et~al.}(2011)\citenamefont {Nomura},
  \citenamefont {Otsuka}, \citenamefont {Shimizu},\ and\ \citenamefont
  {Guo}}]{nomura2011rot}%
  \BibitemOpen
  \bibfield  {author} {\bibinfo {author} {\bibfnamefont {K.}~\bibnamefont
  {Nomura}}, \bibinfo {author} {\bibfnamefont {T.}~\bibnamefont {Otsuka}},
  \bibinfo {author} {\bibfnamefont {N.}~\bibnamefont {Shimizu}}, \ and\
  \bibinfo {author} {\bibfnamefont {L.}~\bibnamefont {Guo}},\ }\href {\doibase
  10.1103/PhysRevC.83.041302} {\bibfield  {journal} {\bibinfo  {journal} {Phys.
  Rev. C}\ }\textbf {\bibinfo {volume} {83}},\ \bibinfo {pages} {041302}
  (\bibinfo {year} {2011})}\BibitemShut {NoStop}%
\bibitem [{\citenamefont {Scholten}(1985)}]{scholten1985}%
  \BibitemOpen
  \bibfield  {author} {\bibinfo {author} {\bibfnamefont {O.}~\bibnamefont
  {Scholten}},\ }\href@noop {} {\bibfield  {journal} {\bibinfo  {journal}
  {Prog. Part. Nucl. Phys.}\ }\textbf {\bibinfo {volume} {14}},\ \bibinfo
  {pages} {189} (\bibinfo {year} {1985})}\BibitemShut {NoStop}%
\bibitem [{\citenamefont {{S. Heinze}}(2008)}]{arbmodel}%
  \BibitemOpen
  \bibfield  {author} {\bibinfo {author} {\bibnamefont {{S. Heinze}}},\
  }\href@noop {} {} (\bibinfo {year} {2008}),\ \bibinfo {note} {computer
  program ARBMODEL (University of Cologne)}\BibitemShut {NoStop}%
\bibitem [{\citenamefont {{Brookhaven National Nuclear Data Center}}()}]{data}%
  \BibitemOpen
  \bibfield  {author} {\bibinfo {author} {\bibnamefont {{Brookhaven National
  Nuclear Data Center}}},\ }\href@noop {} {}\bibinfo {howpublished}
  {{http://www.nndc.bnl.gov}}\BibitemShut {NoStop}%
\bibitem [{\citenamefont {Rzaca-Urban}\ \emph {et~al.}(2013)\citenamefont
  {Rzaca-Urban}, \citenamefont {Urban}, \citenamefont {Smith}, \citenamefont
  {Ahmad},\ and\ \citenamefont {Syntfeld-Ka\ifmmode~\dot{z}\else
  \.{z}\fi{}uch}}]{rzacaurban2013}%
  \BibitemOpen
  \bibfield  {author} {\bibinfo {author} {\bibfnamefont {T.}~\bibnamefont
  {Rzaca-Urban}}, \bibinfo {author} {\bibfnamefont {W.}~\bibnamefont {Urban}},
  \bibinfo {author} {\bibfnamefont {A.~G.}\ \bibnamefont {Smith}}, \bibinfo
  {author} {\bibfnamefont {I.}~\bibnamefont {Ahmad}}, \ and\ \bibinfo {author}
  {\bibfnamefont {A.}~\bibnamefont {Syntfeld-Ka\ifmmode~\dot{z}\else
  \.{z}\fi{}uch}},\ }\href {\doibase 10.1103/PhysRevC.87.031305} {\bibfield
  {journal} {\bibinfo  {journal} {Phys. Rev. C}\ }\textbf {\bibinfo {volume}
  {87}},\ \bibinfo {pages} {031305} (\bibinfo {year} {2013})}\BibitemShut
  {NoStop}%
\end{thebibliography}%

\end{document}